# Observation and Spectral Assignment of a Family of Hexagonal Boron Nitride Lattice Defects

*Daichi Kozawa[1,†], Ananth Govind Rajan[1,†], Sylvia Xin Li[1], Takeo Ichihara[1], Volodymyr B. Koman[1], Yuwen Zeng[1], Matthias Kuehne[1], Satish Kumar Iyemperumal[1], Kevin S. Silmore[1], Dorsa Parviz[1], Pingwei Liu[1,2], Albert Tianxiang Liu[1], Samuel Faucher[1], Zhe Yuan[1], Wenshuo Xu[3], Jamie H. Warner[3], Daniel Blankschtein[1], and Michael S. Strano[1,*]*

[1] Department of Chemical Engineering, Massachusetts Institute of Technology, Cambridge, Massachusetts 02139, United States

[2] College of Chemical and Biological Engineering, Zhejiang University, Hangzhou, 9 Zhejiang Province, China, 310027

[3] Department of Materials, University of Oxford, Parks Road, Oxford, OX1 3PH, United Kingdom

[†]These authors contributed equally

[*]Corresponding Author: Michael S. Strano (E-mail: strano@mit.edu)

Atomic vacancy defects in single unit cell thick hexagonal boron nitride are of significant interest because of their photophysical properties, including single-photon emission, and promising applications in quantum communication and computation. The spectroscopic assignment of emission energies to specific atomic vacancies within the triangular lattice is confounded by the exponential scaling of defect candidates with the number of removed atoms. Herein, we collect more than 1000 spectra consisting of single, isolated zero-phonon lines between 1.69 and 2.25 eV, observing 6 quantized zero-phonon lines arising from hexagonal boron nitride vacancies. A newly developed computational framework for isomer cataloguing significantly narrows the number of candidate vacancies. Direct lattice imaging of hexagonal boron nitride, electronic structure calculations, and subsequent boric acid etching are used to definitively assign the 6 features. Systematic chemical etching supports the assignment by demonstrating the sequence of growth of successively larger vacancy centres from smaller ones, with the defects including a single B vacancy and a 16-atom triangular defect. These features exhibit a range of emission lifetimes from 1 to 6 ns, and phonon sidebands offset by the dominant lattice phonon in hexagonal boron nitride near 1370 cm$^{-1}$. This assignment should significantly advance the solid-state chemistry and photophysics of such vacancy emitters.

Hexagonal boron nitride (hBN) is a layered material that can be exfoliated or chemically synthesized as a single unit cell thick two-dimensional (2D) material, providing molecularly thin dielectrics [1,2], chemically resistant coatings [3,4], and nanoporous membranes [5,6]. There has been recent interest in vacancy centres or atomic defects within the B and N triangular lattice because they demonstrate photoluminescence, similar to diamond nitrogen-vacancy centres [7]. Such defects can exhibit single photon emission [8-13], a property being explored for quantum communication and computation, exploiting the fact that 2D single photon sources are more readily coupled to photonic [14] and optoelectronic components [15]. However, progress towards uncovering the basic physics underlying such vacancy emitters in hBN is hindered by the lack of a structural assignment of emission energies. The task of identifying the structure of an absence of atoms, or an anti-molecule, within a triangular lattice, presents a unique challenge in materials science and photophysics. As the atomic size of the vacancy increases, the number of unique isomers, or atomic configurations, increases exponentially with the number of removed atoms. Indeed, a 10 atom vacancy in a 2D lattice has 444 candidates for assignment (Table S1) [16]. Herein, we present the spectral assignment of six emission lines corresponding to the most common lattice defects that are observable in hBN samples prepared by various methods, including multilayer and monolayer CVD, liquid-phase and mechanical exfoliation. Apart from utilizing a confocal spectroscopic instrument to collect hundreds of single defect spectra, we correlate our results with corresponding scanning transmission electron microscopy (STEM) imaging of the lattice and leverage computational tools for anti-molecule isomer cataloguing and electronic structure calculations. This allows us to assign 6 photoemission zero-phonon lines (ZPLs) to specific hBN vacancies, while ruling out five other hBN vacancies from the assignment. Hence, this assignment should greatly accelerate the basic science and technological applications of hBN.

We first generated and compared more than 1000 photoluminescence spectra from hBN flakes obtained via different preparation methods, using a confocal spectroscopy system (Fig. S1), generating distinct distributions in the emission energy corresponding to ZPLs of vacancy defects. Our first observation is that the emission energies appear to be quantized within approximately 40 meV and are organized into primarily six clusters. For example, liquid exfoliation produces 100 nm hBN flakes well isolated on a $SiO_2$/Si substrate (Fig 1A, B). Confocal (Fig 1C) and hyperspectral (Fig. S2) photoluminescence mapping reveal that some flakes emit distinctive ZPLs over a range of intensities and energies (Fig. 1D). Moreover, the second-order photon correlation function [17] of $P_2$ in liquid-exfoliated hBN shows a clear dip below 0.5 at zero delay time (Fig. 1E), indicating photon antibunching, and confirming single photon emission from the vacancy centre [10]. The deconvolution and binning of approximately 250 ZPLs reveal a histogram organized into 6 peaks at 2.25, 2.15, 1.98, 1.90, 1.81, and 1.69 eV that we label $P_1$, $P_2$, $P_3$, $P_4$, $P_5$ and $P_6$, respectively. The relative probabilities of these Gaussian peaks observed in the hBN flakes are notably altered with sample preparation, from exfoliation (Fig 1F), to 3 nm thick multilayer CVD (Fig 1G and Fig. S4), and monolayer CVD samples (Fig 1H). However, the discretization of the emission energies and the peak positions appear notably invariant. The 57±21 meV line width of these distributions in the histogram is much broader than the 10 meV of a typical ZPL. However, the former value approximately coincides with the observed temporal, stochastic fluctuations in the emission energy that we observe for $P_1$ to $P_6$ and discuss later in this article. The samples exhibit strong phonon modes in their Raman spectra at wavenumbers of 1367.5, 1368.2, and 1369.5 cm$^{-1}$ (Fig. S5) for liquid-exfoliated, multilayer CVD, and monolayer CVD samples, respectively, consistent with previous reports [18]. The characteristic Lorentzian line shape of the ZPL [19] assigns these features to embedded vacancies within the hBN lattice. The above observations reinforce the quantized nature of these features. Normalizing the populations on an areal basis reveals greater numbers of vacancies for liquid-exfoliated samples (0.460/μm$^2$, Fig 1I), compared to multilayer (0.0579/μm$^2$, Fig 1J) and monolayer (0.0238/μm$^2$, Fig 1K) CVD samples.

The ZPL emission from the hBN samples arises due to intrinsic defects in the form of lattice vacancies corresponding to the removal of integer numbers of atoms from the pristine, triangular hBN lattice [8]. The problem of linking these six quantized features $P_1$ to $P_6$ to specific vacancy defects in the hBN triangular lattice is confounded by the exponential number of candidate *polyiamonds*, or isomeric shapes of defects [16], with the number of lattice atoms removed (*N*). For example, a monoatomic triangular lattice, such as, graphene, with removed atoms *N* = 1, 9, and 16, yields 1, 159, and 189309 as the number of unique isomers for lattice vacancies, respectively [16], with larger numbers for the diatomic hBN unit cell. We have recently formulated a computational framework to solve what we call the *Isomer Cataloguing Problem*, to narrow down the vacancy defect candidates into the unique, most-probable ones in 2D lattices [20]. The framework combines electronic structure calculations, kinetic Monte Carlo (KMC) simulations, and chemical graph theory consistent with the underlying symmetries of 2D materials, to generate a catalogue of unique, most-probable isomers of 2D lattice defects, and to predict their rates of inter-conversion [20]. For graphene, a 2D triangular lattice material similar to hBN, our solution to the *Isomer Cataloguing Problem* has been shown to predict exact defect morphologies in the carbon lattice corroborated using STEM imaging, including capturing the dynamics of isomerization [21]. In the case of hBN, the approach predicts just one stable, low energy defect, as opposed to multiple defects in the case of graphene, for a given number of atoms removed, $N$. Comparative reaction kinetics shows that boron (B) atoms etch much faster than nitrogen (N) atoms [20], consistent with experimental reports that lattice defects in hBN appear invariably terminated with N atoms [22-26]. When $N$ is a perfect square number of atoms removed ($N$ = 1, 4, 9, 16, …), the energetically most favourable defect is a perfect triangle and the *Isomer Cataloguing* approach predicts this (Table S2), in agreement with several STEM imaging studies [22,23,27,28]. The fact that our algorithm predicts just one stable defect structure for all values of $N$ in the case of hBN greatly simplifies the number of possible assignment candidates for $P_1$ through $P_6$.

We utilized electronic structure density functional theory (DFT) calculations to determine the relative electronic energy scaling of the 16 smallest, most-probable lattice defects in hBN ($N$=1 through 16). We also considered transformations of these defects into slightly different variants [32], due to atomic rearrangements in the lattice. For this purpose, we calculated the activation barriers for edge hopping of boron and nitrogen atoms in the hBN lattice, along the pore mouth. As in the graphene case [20], most of the edge hopping events have significantly large activation barriers due to the hopping atoms being uncoordinated (Fig. S8-11). Considering only those edge hopping events with barriers less than 1.2 eV, 6 additional defects, which are kinetically reachable from the intrinsic defects in hBN, are added to the set and are labelled with an asterisk (*). These defects consist of a nitrogen atom occupying an interstitial boron position, leading to the formation of an N–N–N chain. This leads to a total of 22 low energy defect structures forming candidates for the assignment (Table S6).

Hybrid DFT calculations were used to determine the energy of the highest occupied electronic level (HOEL) and the lowest unoccupied electronic level (LUEL) for each defect isomer identified above. Note that, more accurate calculation schemes, such as the GW method [33], and the Bethe-Salpeter equation[34], are available for simulation of the optical properties of materials [35], including hBN [36]; however, these methods at present are too computationally expensive to be carried out for 22 different defect systems, as we have done in this study. Indeed, many authors have successfully utilized hybrid DFT for describing hBN systems [37-39], to ensure a balance between computational expense and accuracy. Accordingly, we utilized hybrid DFT [40] for the HOEL-LUEL gap calculations because it incorporates a fraction of exact exchange (Pauli's exclusion principle) in the description of the system [40], and therefore, is able to predict band gaps and defect energy levels in better agreement with the experimental data [41]. For example, the band gap of pristine hBN is predicted to be 4.7 eV by standard generalized gradient approximation (GGA) DFT, while hybrid DFT predicts the band gap of pristine hBN to be 5.8 eV, in closer agreement with the experimental value [42] of 5.955 eV (Fig. S12). Applying hybrid DFT calculations to the set of the 16 most-probable defects identified above (Fig. S13) yields the plot depicted in Fig. 2A for the predicted HOEL-LUEL electronic gap (Fig. S14-20) as a function of the defect size ($N$) in monolayer hBN up to $N$=16. The HOEL-LUEL gap predicted using hybrid DFT is known to directly correlate with the first emission energy of a given electronic system [43]. Indeed, the calculated gaps for the set of hBN defects fall within 1.6 to 2.4 eV, in quantitative agreement with the observed emission energies $P_1$ though $P_6$. Conversely, a non-hybrid, GGA DFT calculation predicts energy gaps of less than 0.5 eV. When projected onto our observed spectral range of 1.6 to 2.4 eV, the hybrid DFT calculations predict 6 features spaced approximately 0.1 eV in energy, in quantitative agreement with the experimental observations. Our hybrid DFT calculations exclude several candidate defects from the emission energy assignment. Specifically, our calculations rule out defects $N$=2 and $N$=4, because they have very large simulated HOEL-LUEL gaps (beyond 4 eV), as well as defects $N$=3, 11, and 13, because they have very small HOEL-LUEL gaps (less than 1.5 eV), that are outside of our observation range.

The set of the kinetically stable defect isomers, and the relative scaling of the HOEL-LUEL gap energies yield only two possible candidate assignments of specific lattice defects to experimental emission energy peaks from $P_1$ through $P_6$. Fig. 2B shows these two assignments (i) and (ii) as parity plots of calculated DFT electronic gap energies plotted versus the observed ZPL emission energies (Tables S7). Both show near parity agreement with an offset linear trend but can be readily differentiated by the $N$=1 (1,0) assignment as $P_3$ in (i) and as $P_2$ in (ii). Note that ($N_B$, $N_N$) denotes the number of B and N atoms removed, such that $N = N_B + N_N$. Using STEM imaging, we experimentally observe that the monolayer hBN samples exhibit 62.5% of imaged defects (10 of 16) as $N$=1 (1,0), and the multilayer (bilayer region) hBN samples exhibit 59.1% of imaged defects (13 of 40) as $N$=1 (1,0), corresponding to the B monovacancy (Fig. 2C and Fig. S22A). Recall that these monolayer hBN samples specifically show a clear mode in the emission distribution at 1.95 eV, corresponding to the $P_3$ feature (Fig. 1G and 1J). This anchor point clearly supports the first assignment (i) over (ii). As additional evidence supporting (i), We find 36.3% of imaged defects (8 of 22: 1 of $N$=1, 3 of $N$=1, 4 of $N$=16) as triangular defects including $N$=8 (5,3), $N$=9 (6,3), $N$=16 (10,6) assigned to be $P_2$ (Fig. 2C and Fig. S22B, Fig. S23, Table S8). In fact, further support for assignment (i) is provided by the observation that $P_2$ is abundant in liquid-exfoliated and multilayer CVD samples, in agreement with its assignment to highly stable perfect triangular defects ($N$=9, 16) (Table S2) in assignment (i).

Vacancy-impurities such as C, O, and Si have been previously suggested as components of some colour centres in hBN [29-31]. However, we observe that $P_1$ through $P_6$ appear regardless of sample preparation from liquid exfoliation to CVD growth (Fig. 1F-H). Also, we find that 87 ZPLs observed in 19 literature (Fig. S7) studies share the same discretization that we observe. Hence, we rule out heteroatom insertions into the triangular lattice defects as possibilities for our assignment of the emission energies. Moreover, as shown later, the assignment of peak $P_3$ is independent our calculations or chemical etching experiments; instead it follows directly from our experimental imaging data. Furthermore, the visualization of abundant monovacancy in the STEM images clarifies that no heteroatom configuration is available. The STEM images show no signature of heteroatoms, where heavier atoms such as Si exhibit higher contrast in STEM.

We note that the observed photoluminescence includes several effects such as the many body interactions of electrons and holes, *i.e.*, the excitonic effect in hBN [42], which leads to the formation of defect bound excitons [36]. As mentioned before, the solution of the Bethe-Salpeter equation with GW-correlated quasiparticle energies [48] predicts more accurate electronic gaps. The spectral shape and energy are subject to Jahn-Teller effect on exciton states [49], vibronic coupling, and the optical selection rules that result from defects having different symmetries. However, we emphasize that we do not rely solely on the DFT calculations, but also on the STEM images and chemical etching by boric acid to independently confirm the assignment, culminating in a combined experimental-theoretical assignment. Hybrid DFT calculations are only used in this work to determine the *qualitative* rank ordering of the energy of the HOEL-LUEL gap for each defect isomer. The most frequent observation of monovacancy in the STEM of monolayer hBN plays the role of an anchor point for peak $P_3$, which is observed in abundance in the same monolayer hBN. Boric acid converts smaller defects of $P_3$ ($N$=1), $P_4$ ($N$=5, 7), $P_5$ ($N$=6) into larger defects of $P_2$ ($N$=9, 16). The expansion of the smaller defects is kinetically trapped at the triangular defects of $P_2$ ($N$=9, 16), again suggesting agreement between our theoretical findings and experimental results.

We further tested the assignment by deliberate perturbation of the hBN defect distribution, using boric acid to etch existing defects in the lattice. Boric acid forms hydroxyl radicals (•OH) at high temperatures [44], thereby promoting the etching [45] of dangling bonds [46] in the lattice vacancies. Boric acid at high temperatures is known to form adsorbed OH groups on graphene, functioning as a lattice etchant [47]. The increase in the number of atoms removed ($N$) due to etching should specifically convert smaller defects into larger ones, allowing for a clear test of assignment (i). We annealed liquid-exfoliated hBN flakes deposited on $SiO_2$/Si substrates with solid boric acid powder at 850 °C for 30 min, and then measured the resulting photoluminescence spectra. Remarkably, we observe that populations of $P_3$, $P_4$, $P_5$ and $P_6$ decrease, while $P_2$ clearly increases (Fig. 3A) consistent with exclusively smaller and non-triangular $N$=5, 7, 6 vacancy defects increasing to larger and stable triangular versions ($N$=9, 16) due to etching by boric acid. Boric acid treatment of monolayer CVD hBN further supports assignment (i). In this case, the etching diminishes $P_3$ ($N$=1), producing more $P_2$, and also generating the lower energy emission labelled as $P_6$ (Fig. 3B), a peak which is assigned to the $N$=12 vacancy, and is observed rarely in liquid-exfoliated and multilayer CVD hBN, but is not seen in monolayer CVD hBN (Fig. 1F-K). Noted that thermal annealing alone, and a wide range of other processing conditions, do not alter the defect population distribution (Fig. S28). These experiments support the relative size scaling of the defects proposed in assignment (i), because, as expected, the etching produces inter-conversions from smaller to larger $N$. They also support the $N$=9 and 16 assignments as chemically stable triangular lattice defects.

As a stringent test of the specific ordering of the assignment and to generate additional supporting evidence, we carried out the systematic boric acid etching of liquid-exfoliated hBN with a lower etching temperature and shorter etching time shown in Fig. S29. We extracted the monovacancy defect density from $P_3$ ZPLs in hyperspectral maps with various degrees of etching. We observe the onset of the $P_3$ defect density at 650 °C, which reaches equilibrium at 850 °C of annealing temperature. A first-order reaction model for the monovacancy density describes the observed temperature and time dependence of the etching reaction (detailed derivation in the Supplementary Information). We extracted the reaction rate constants of $k_a$ = 0.0288 min$^{-1}$, $k_b$ = 0.125 min$^{-1}$ by fitting the temporal data (see Supplementary Note "**Reaction Analysis of Boric Acid Etching**"). We also found that the activation energy for etching of triangular defects, $E_b$ = 2.30 eV matches the activation energy predicted using the *Isomer Cataloguing* Framework (Table S2).

The transitions from smaller to larger defect sizes allow us to determine the correctness of the assignment. Figure S29 shows the defect density of each ZPL peak as a function of the reaction time with boric acid at 850 °C. To describe the observed reaction kinetics, we introduce a hidden Markov model (Fig. 3C). We assume that the first-order reactions by which the state $N_i$ transitions to the state $N_{i+1}$ are irreversible. The hidden states consist of $N$=1 through 16 and next to the range $N$=0 and 17 at the reaction rate of $k_a$, $k_b$, and $k_c$ for etching hBN lattice, non-triangular and triangular defects, respectively. We can only observe the emission states as ZPLs assigned to $P_1$ through $P_6$ and all other $N$s are assigned to un-observable peaks labeled as $P_{dark}$. Fitting the reaction model to the time variation of the experimental defect densities, we extracted the rate constants $k_c$ = 0.267 min$^{-1}$ at 850 °C (Figure S30) and $k_c$ = 0.00868 min$^{-1}$ at 220 °C (Fig. S29).

We next compare the RMS error of the kinetic model fit for our assignment with the respective RMS errors of kinetic model fits for several randomly made assignments of the defects to the emission peaks. First, we assign the 18 vacancy states ($N_0$ to $N_{17}$) to the 7 groups ($P_1$ to $P_6$, $P_{dark}$) randomly (Fig. 3C). Each vacancy state $N_i$ is uniformly assigned to either of the 7 groups (*i.e.,* the probability of $N_i$ being assigned to $P_j$ is the same). Among these random assignments, we only count the assignments where each group $P_j$ has at least one vacancy state $N_i$ assigned to it. To avoid overfitting, we use only 3 different $k_i$ to denote the reaction constants $k_a$, $k_b$ and $k_c$ used above. For this random assignment approach, the total number of permutations will be more than $7^{18}$-$7\times6^{18}$. Instead of testing all the permutations, we checked 8112 random assignments out of them (Fig. 3D). We found that only 0.7% of the 8112 random assignments have smaller RMS error than our assignment. Among these, only 10 have $N_0$ assigned to $P_{dark}$. Since $N_0$ represents the pristine lattice, it should not be grouped with the other peaks. We use $P_3$ as an anchor, and find only one case among the above 10 assignments where $N_0$ is assigned to $P_{dark}$ and $N_1$ is assigned to $P_3$. For this case, however, $k_a$ is found to be 3.68×10$^{-34}$ min$^{-1}$, which is essentially zero, and the assignment is therefore unphysical. Although we did not run all the possibilities, the emergent statistics corroborate that our assignment is among the optimal choices.

Based on our previously introduced *Isomer Cataloguing* framework[20], we have 5 first-order rate constants for the etching process: $k_d$, $k_e$ for etching rate of non-triangular defects in addition to $k_a$, $k_b$, $k_c$ (see Fig. 3C, details are in the Supplementary Note "**Reaction Analysis of Boric Acid Etching**", based on the data in Table S2). Fixing $k_a$, $k_b$, $k_c$ based on the three-parameter model above, we varied $k_d$ and $k_e$ and fitted the 5-parameter model to the kinetic experiments (Fig. 3E). The reaction model describes the distribution and the reaction time dependence of $P_1$ through $P_6$ and $P_{dark}$. Therefore, the etching results strongly support our assignment without relying even on the *qualitative* accuracy of the calculated electronic gap energies.

Photoluminescence excitation spectroscopy [50,51] provides insight into the electronic structures of $P_1$ through $P_6$. We measured photoluminescence excitation spectra of the ZPLs in liquid-exfoliated hBN because we observe an absence of blinking for this type of sample (Fig. S31). The excitation spectra for $P_1$ through $P_6$ (Fig. 4A, B) reveal additional peaks that are notably broader than the ZPLs. The additional broad peaks in the excitation spectra are assigned to phonon side bands [32], despite missing ZPLs in the excitation spectra, an artefact caused by the necessary cut off wavelength of long pass filters. Note that the energy difference between the ZPLs and the assigned phonon side bands for $P_1$ through $P_5$ are between 158 to 166 meV ($P_6$ side bands are missing), as tabulated in Table S10. These energy differences between the ZPLs for $P_1$ through $P_5$ correspond to phonon energies of 1274 to 1339 cm$^{-1}$ in wavenumber (Fig. 4C), comparable approximately to the dominant lattice phonon in hBN [12] observed in the Raman spectrum at 1367.5 cm$^{-1}$. Note that lattice phonons can augment or subtract from the ZPL energy, resulting in a symmetric and discretized distribution of additional bands about the ZPL. While the ZPL is approximately Lorentzian (+), the side bands are Gaussian peaks due to one- (*) or two- (^) phonon interactions (Fig. S32).

It is also useful to examine the photoluminescence lifetime of liquid-exfoliated hBN because, again, the 66 nm thick flakes from this sample exhibit no blinking over 300 s (Fig. S31). A typical photoluminescence decay curve for $P_2$ is shown in Fig. S33A, described by a single exponential decay. Interestingly, we do not observe a dependence of the photoluminescence lifetime on the assigned lattice structure of the emitters for $P_1$ through $P_6$ (Fig. 4D). Note that the photoluminescence lifetime typically reflects the dielectric environment of the photo-emitters. This influence is amplified in atomically thin materials because the structures are likely exposed to the external environment [52]. Defects close to the surface of the hBN crystal matrix or to the supporting substrate can lose energy to external gas or solid energy sinks, necessarily resulting in shorter lifetimes (Fig. S33B). Conversely, a defect deep within the hBN crystal should exhibit longer lifetimes. Hence, the observed scatter of lifetimes in Fig. 4D is attributed to the random distribution of vacancies within the crystal. As a test, we measured the lifetimes of $P_1$ through $P_4$ in monolayer CVD hBN, which necessarily must reside in contact with the substrate and the external atmosphere. As expected, Fig. 4D shows that the lifetimes in this case are uniformly three times shorter.

We used our assignment to further elucidate the photophysical properties of the vacancies to which $P_1$ through $P_6$ are assigned. Interestingly, an excitation polarization measurement of liquid-exfoliated hBN shows two-fold symmetry for all of the assignments $P_1$ through $P_6$ (Fig. 5A), as reported before for similar defects [19], where B and N atoms on the edge of the lattice vacancy form in-plane dipoles [53]. While $P_3$ as a mono-vacancy exhibits 3-fold structural symmetry, we observe that the photoexcited electron breaks this symmetry into a dipole [53]. Time-resolved polarization measurements show an even more nuanced picture. The photoluminescence of multilayer CVD hBN (Fig. 5B) was measured using a setup where the optical configuration (Fig. S34) allowed us to rotate the linear polarization of the laser light by 25 deg/s and collect spectra every 250 ms. We observe blinking characterized by stochastic fluctuations in both emission intensity and wavelength, as reported earlier [37,54], similar to quantum dots [55] and NV centres of nanodiamonds [56]. This blinking and spectral fluctuation of the ZPLs of multilayer CVD hBN are observed for all assigned features $P_1$ through $P_6$, with the polarization switching its dipole orientation stochastically. For all features, we observe two or three discrete blinking states in the time traces (Fig. 5C), with corresponding dipole switching (Fig. 5D). A stable electric dipole of a vacancy centre at a fixed polarization angle should exhibit constant oscillation, resulting in an angular dependent intensity given by:

$$I(\theta) = I_0 \cos^2 \theta \qquad \text{(Equation 1)}$$

where $I$ is the integrated intensity of a peak, $\theta$ is the angle of linear polarization of the laser, and $I_0$ is the maximum intensity over the time period. We fitted the intensity trajectories by introducing a phase shift $\varphi$ in the polarization angle of the electric dipole as follows:

$$I(\theta, \varphi) = I_0 \cos^2(\theta + \varphi) \qquad \text{(Equation 2)}$$

where $\varphi$ was determined at each period of $\cos^2 \theta$, as shown in Fig. 5C. The photoluminescence intensity and emission energy at each end of the period were also extracted (Fig. S35-S36). The extracted time traces show discrete states (Fig. 5D). To fit the time traces of the intensity, the polarization angle, and the emission energy, the most plausible numbers of states were estimated using the silhouette method [57], and the centroids of the states were determined using the Gaussian mixture model [58], both of which are typical machine learning methods used to evaluate clusters. We tentatively assign the cause of switching in all cases to be extrinsic because the distributions of the dwell times in intensity, phase angle, and energy all show exponential distributions (Fig. S37), with dynamics independent of the assignment. Indeed, time constants of the distributions span 4.0 to 11.8 s, independent of the assignment (Table S11). This curious switching, which seems to be independent of the assigned vacancy structure, is consistent with molecular or adatom adsorption to the defect, with relatively independent adsorption energies for each structure.

To further test the mechanism of defect blinking, we continuously acquired photoluminescence spectra over 150 s, and compared the spectral trajectories for multilayer CVD hBN samples on $SiO_2$/Si substrates with different passivation methods applied, including the as-transferred sample (Fig. 5E-i), a $N_2$ purged sample (E-ii), a graphene-covered sample (E-iii), a PMMA-covered sample (E-iv), and a sample with an $Al_2O_3$ atomic layer deposited underneath hBN (E-v). These passivation methods can prevent external gas adsorption, including $O_2$ and $H_2O$, from the atmosphere. Surprisingly, these five systems show blinking, as characterized by both intensity and wavelength fluctuations, regardless of the passivation method. Ambient $O_2$ is a common quencher of many molecular fluorescent emitters [59], but is readily excluded using an $N_2$ purge, or a graphene or PMMA overcoating film (E-ii to E-iv). Likewise, charged impurities in $SiO_2$ also quench fluorescence in carbon nanotubes [48], but this plausible quenching is mitigated using an intervening insulating layer, such as, $Al_2O_3$ [60]. Therefore, we rule out external molecular adsorption, or the substrate, as the sources of the persistent blinking of hBN defects. This suggests that internally trapped molecules or adatoms, within the crystal matrix itself, and around the vacancies, may be responsible for the stochastic blinking. In one particular scenario, spectral fluctuations may be caused by the hopping of freed hBN lattice atoms along the lattice vacancies, thereby altering the polarization angle and the emission energy of the electric dipole due to the transient structural modulation. To investigate whether connected lattice atoms can participate in this mechanism, we calculated hopping barriers required to transform a vacancy from the initial, most-probable state to several final configurations, using our *Isomer Cataloguing* framework (Fig. S8-11), and estimated the resulting time constants associated with the switching or potential blinking (Fig. S38) of defects. Our calculations show that only the time constants of $P_2$, $P_4$, and $P_5$ are in the range between 4 and 12 s at a system temperature of 100-150 °C, with the other peaks being excluded, which is inconsistent with a lattice rearrangement of B and N atoms being the operative mechanism for the blinking of defects.

The structural assignment of hBN defects should prove useful for identifying processes that interconvert known vacancies to new defect structures. We applied probe tip ultra-sonication sonication to liquid-exfoliated hBN, followed by annealing at 850 °C, to activate the vacancy centres. Tip sonication significantly modifies the emission of liquid-exfoliated hBN. The median thickness and lateral size of the liquid-exfoliated hBN are found to be 66 nm (10th and 90th percentiles: 39 and 129 nm) and 358 nm (58 and 1013 nm). Tip sonication for 5 min generates thinner and smaller flakes having a median thickness and lateral size of 9 nm (3 and 25 nm) and 97 nm (48 and 376 nm) (Fig. S39). No apparent difference is observed in X-ray photoelectron spectroscopy before and after the tip-sonication (Fig. S6 and S40). However, the photoluminescence excitation map of the tip sonicated hBN shows a new, sharp emission at 1.785 eV (Fig. 5F), labelled $P_7$, near but measurably smaller than our assigned $P_5$ by 50 meV, and dominant in the new defect distribution. We excluded $P_7$ among the assigned vacancy centres because the evidence is consistent with its formation along the perimeter of the hBN flake. The angularly dependent photoluminescence of the $P_7$ peak exhibits an isotropic polarization dependence (Fig. 5H). The strength of the emission indicates multiple P7 defects per flake and we propose that tip sonication forms the $P_7$ vacancy centres along the edge (Fig. 5I), resulting in the superposition of multiple mirror symmetric polarizations, and resulting in the observed isotropic angular dependence. The acoustic cavitation from micro-bubble collapse is known to generate a destructive water jet [61] as well as temperatures reaching 5000 °C, with defect generation seen in carbon nanotubes [62]. We find that the probability to observe the $P_7$ peak increases up to 50% with ultra-sonication time after 60 min (Fig. 5G). However, its association with the edge complicates the assignment of the structure. We include the observation of $P_7$ here to underscore the utility of our spectral assignment to further identify and classify new defects, even those close in energy to the set of six introduced above.

In conclusion, we are able to map each of the six quantized emission energies in hBN samples, prepared using a wide range of synthetic methods, to theoretically-predicted lattice vacancy defects in hBN. Our results provide a key structure-property relationship for such vacancy defects and promise spatial control of emission wavelengths through the atomistic design of defect shapes.

**Author Contributions**

**Acknowledgments**

This work was funded by the Army Research Office via award no. 64655-CH-ISN to the Institute for Soldier Nanotechnologies. The authors acknowledge support from 2015 US Office of Naval Research Multi University Research Initiative (MURI) grant on Foldable and Adaptive Two-Dimensional Electronics (FATE) at MIT, Harvard University and University of Southern California. We appreciate characterization support from The MIT Center for Materials Science and Engineering. D.K. is supported by the Grant-in-Aid for JSPS Fellows (JSPS KAKENHI Grant Number 15J07423) and Encouragement of Young Scientists (B) (JSPS KAKENHI Grant Number JP16K17485) from Japan Society for the Promotion of Science. D.B. and A.G.R. acknowledge the National Science Foundation (CBET-1511526). This work used supercomputing resources provided by the Extreme Science and Engineering Discovery Environment (XSEDE), which is supported by the National Science Foundation. V.B.K. is supported by The Swiss National Science Foundation (projects no. P2ELP3_162149 and P300P2_174469). K.S.S. is supported by the Department of Energy Computational Science Graduate Fellowship program under grant DE-FG02-97ER25308. Microfabrication for this work was performed at the Harvard University Center for Nanoscale Systems (CNS), a member of the National Nanotechnology Coordinated Infrastructure Network (NNCI), which is supported by the National Science Foundation under NSF award no. 1541959.

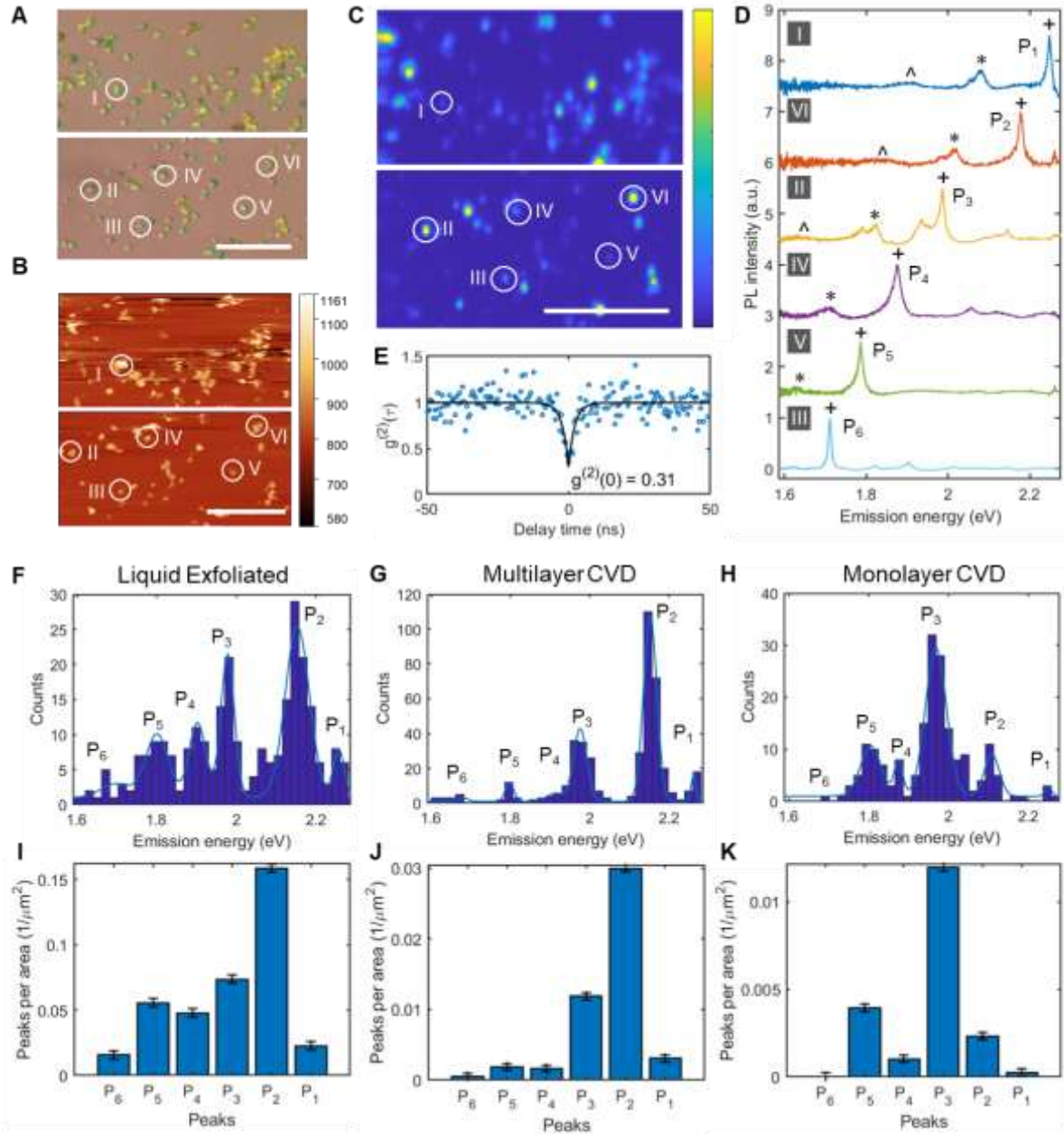

**Figure 1: Observation of discrete emission energies in hBN.** Optical micrograph (**A**) and atomic force micrograph (**B**) showing 100 nm thick hBN flakes from liquid phase exfoliation. Confocal photoluminescence imaging (**C**) reveals emission corresponding to vacancy defects in the lattice showing characteristic zero-phonon lines between 1.6 and 2.2 eV as represented by select spectra (**D**) for isolated defects indicated by the labels of I through VI as shown (scale bar = 10 μm). The spectra consist of zero-phonon line (+), one-phonon (*) and two-phonon (^) sidebands. The photon antibunching curve (**E**) of $P_2$ in liquid exfoliated hBN with a fit using a two-level model and is normalized, where the dip below 0.5 at zero delay time indicates that the vacancy center exhibits single photon emission. Liquid exfoliated hBN

produces a histogram (**F**) of quantized emission energies (labeled $P_1$ through $P_6$) with modes at 2.25 ($P_1$), 2.15 ($P_2$), 1.98 ($P_3$), 1.90 ($P_4$), 1.81 ($P_5$), and 1.69 ($P_6$) eV, where a peak finding algorithm applied to spectra extracted energies of zero-phonon lines. The bin width of 20 meV for the histogram corresponds to the range of temporal fluctuations in energy. Comparison with histograms for multilayer (**G**) and monolayer (**H**) grown by chemical vapor deposition (CVD) show different distributions of the same quantized peaks, where liquid exfoliated hBN (**I**) and multilayer CVD hBN (**J**) hosts more $P_2$ following $P_3$, while monolayer CVD hBN (**K**) hosts abundant $P_3$, suggesting that these abundances reflect structural preferences dependent on the preparation method. Bar charts represent Gaussian curves fit to local maxima in the histogram, normalized by the total scanned area of hBN crystals. Error bars are estimated as the standard deviation of the regression.

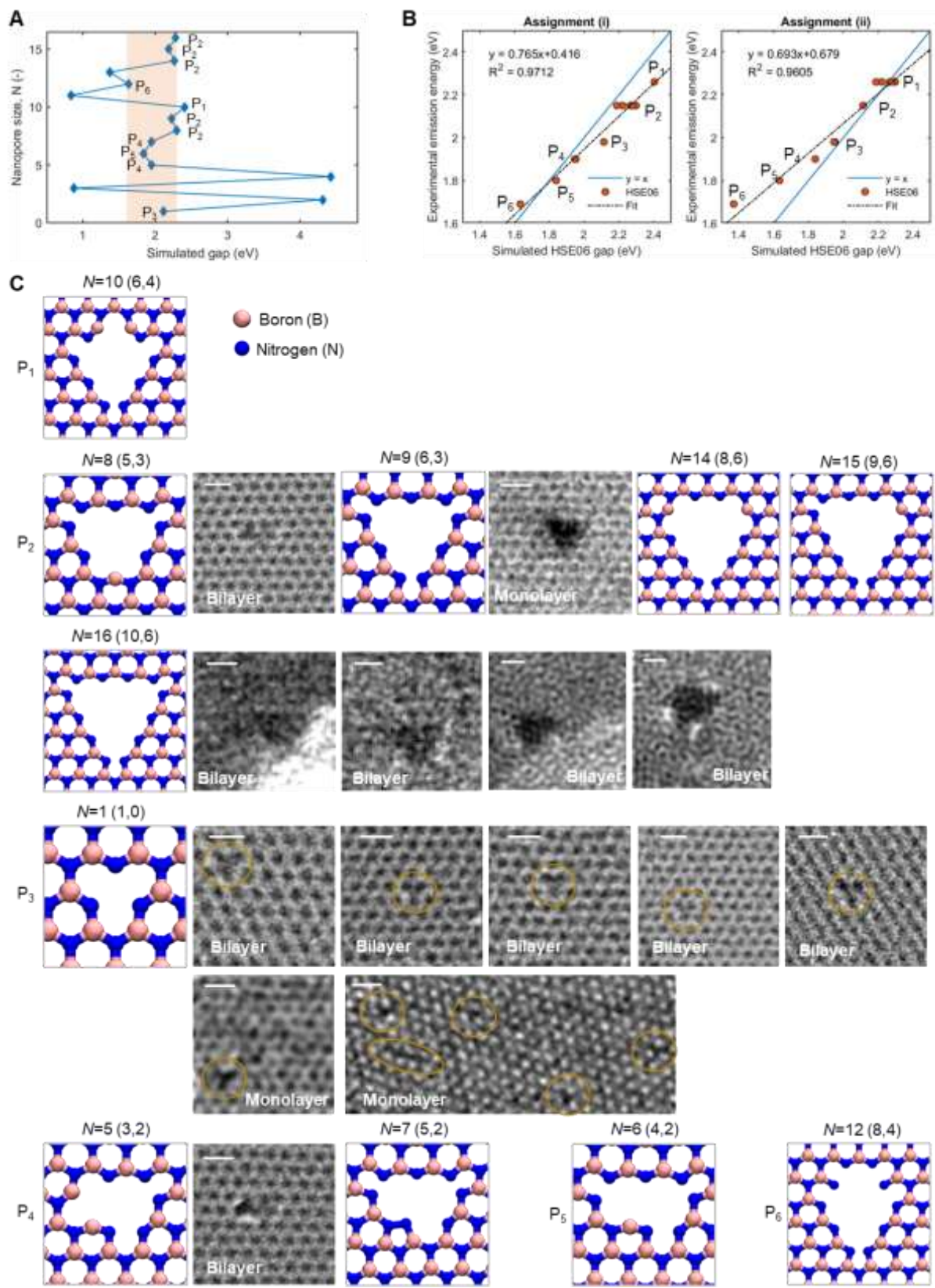

**Figure 2: Spectral assignment of a family of hexagonal boron nitride lattice defects.** (**A**) Electronic energy gaps for the most-probable isomers of defects in hBN, calculated using density functional theory (DFT), with the DFT simulated gaps on the *x*-axis and the defect

sizes on the *y*-axis. The orange band denotes the spectral range over 1.6 through 2.3 eV examined experimentally in this work. Each defect size is assigned to an experimental peak from $P_1$ through $P_6$, based on a descending rank ordering of the simulated electronic energy gaps. (**B**) Two different parity plots comparing the experimental emission energies and the simulated DFT gap energies, based on two ways of assigning defect sizes to experimental emission energy peaks from $P_1$ through $P_6$. In the left panel, Assignment (i), the simulated energy gap of 2.404 eV ($N$=10) is assigned to $P_1$, whereas in the right panel, Assignment (ii), the simulated energy gaps of 2.298 eV through 2.188 eV ($N$=8, 16, 14, 9, 15) are assigned to peak $P_1$. In both cases, the remaining energy gaps are assigned based on a descending rank ordering of energies. In an assignment (i), triangular defects ($N$=9, 16) get assigned to peak $P_2$, which agrees with the high probabilities of triangular defects predicted by KMC simulations and that of peak $P_2$ in the experimentally measured emission spectrum. In assignment (ii), triangular defects ($N$=9, 16) get assigned to peak $P_1$, which attributes high probability triangular defects to a low probability peak ($P_1$). (**C**) Atomic models of the most-probable isomer structures of hBN defects for the assigned peaks, which have been energetically optimized based on DFT calculations. The corresponding STEM images of defects are placed on the right, wherein the scale bars correspond to 5 Å and circles indicate monovacancies. STEM images of several monovacancies ($N$=1) and triangular defect ($N$=9 and 16) are compared.

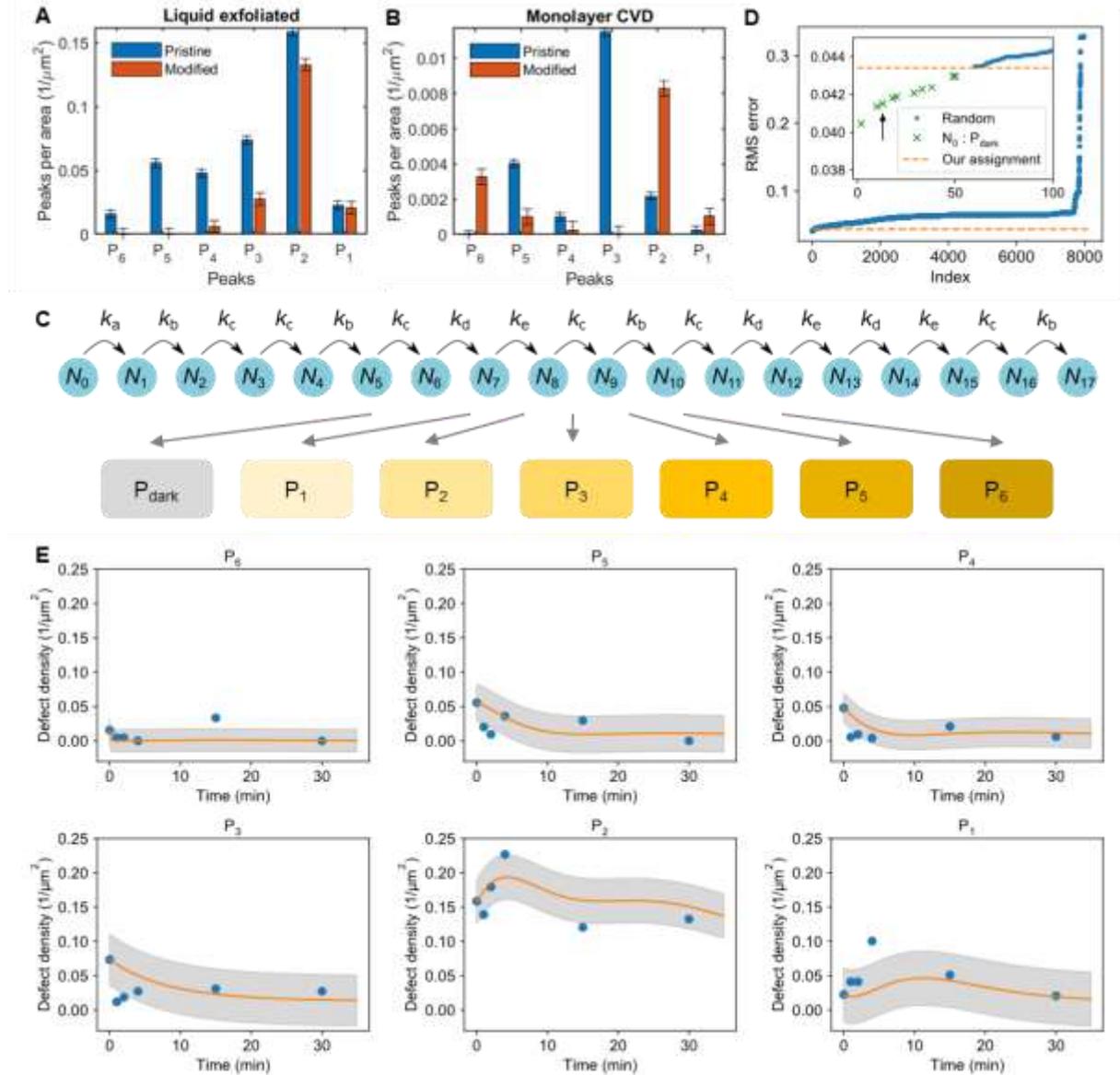

**Figure 3: Modification of defects distribution by boric acid etching.** Bar charts of the frequency of emission peaks per unit area for (**A**) liquid-phase exfoliated hBN and (**B**) monolayer CVD hBN, as is (blue), and after reaction with a boric acid (orange) at 850 °C of 30 min hold time. The probability of peak $P_2$ in the hBN sample, assigned to triangular defects, increases upon boric acid treatment, consistent with the conversion of smaller defects into larger, more stable triangular defects from etching. Error bars correspond to the standard deviation of the regression. (**C**) The transition from a pristine lattice ($N$=0) to $N$=17 (or $N$>16) lattice vacancy can be represented using five first order reaction rate constants $k_a$, $k_b$, $k_c$, $k_d$, and $k_e$ that represent the etching of the lattice. The categorization of the rate constants is based on the calculated lifetimes for each defect size (Table S2). ZPLs outside the detection range were assigned to $P_{dark}$. (**D**) Root mean square (RMS) error, out of 8112 random three

fitting parameters applied to the boric acid reaction kinetics. The RMS error was sorted in an ascending order. Our assignment ranks at 0.7 percentile shown as the orange broken line, validating the accuracy of the assignment. The inset displays the zoom-in around the assignment. The arrow indicates the only one assignment, that has both $N_0$ to be $P_{dark}$ and $N_1$ to be $P_3$, but for that assignment, $k_a \sim 10^{-34}$ min$^{-1}$, which is an unphysical rate constant at 850 °C. (**E**) Assigned defect density with annealing time at 850 °C. The model based on the numerical solution of rate equations describes the distribution of the assignments, showing the reaction rate constants of $k_a = 0.00287$, $k_b = 0.125$, $k_c = 0.267$, $k_d = 0.287$, and $k_e = 9.64$ min$^{-1}$. The errors were estimated from the standard deviation between the experiments and fits.

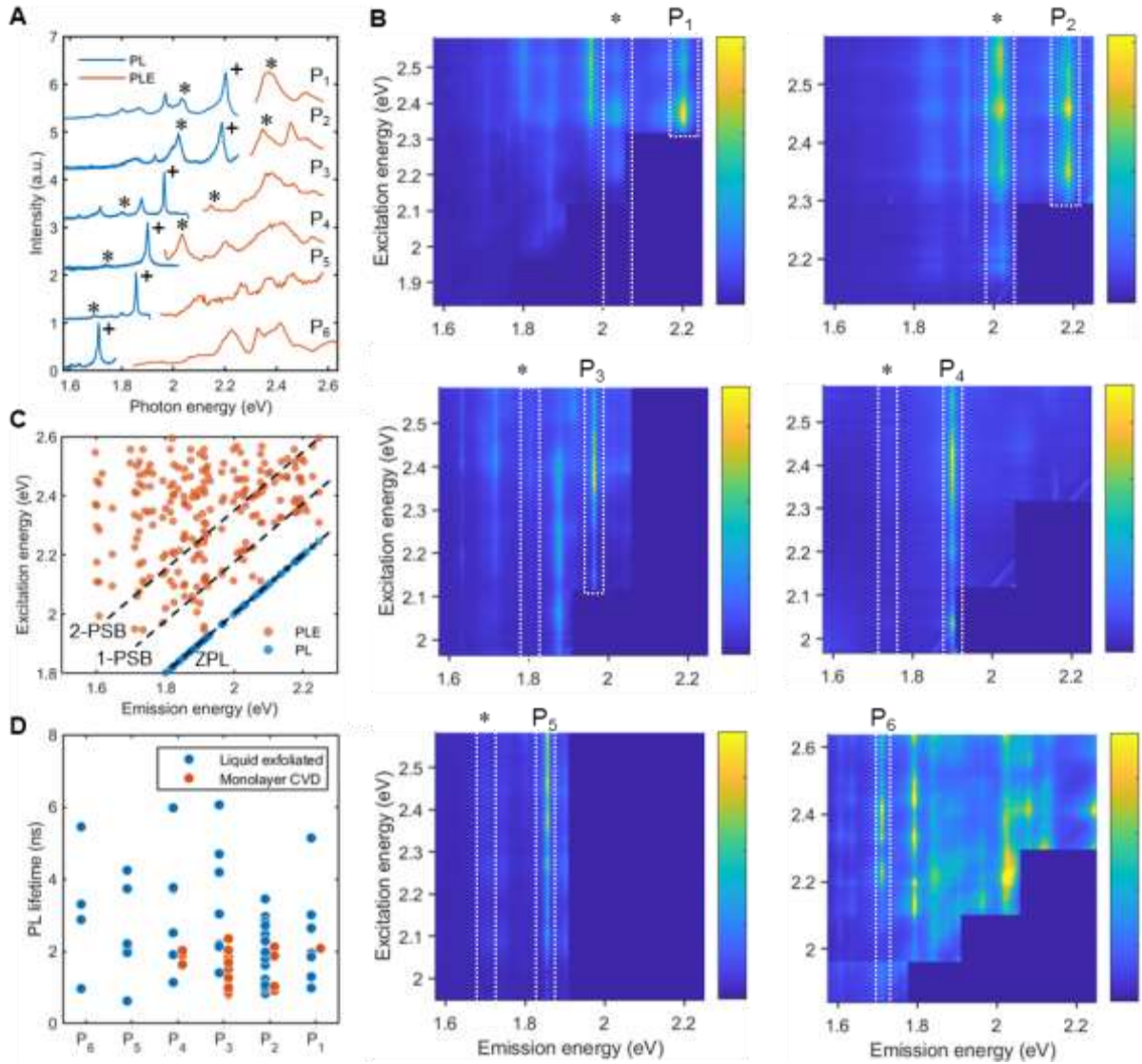

**Figure 4: Individual optical properties of $P_1$ through $P_6$.** Photoluminescence (PL) and excitation (PLE) spectra (**A**) of liquid exfoliated hBN for $P_1$ through $P_6$ showing different resonant energy to excitation, where the zero-phonon line and phonon sideband are labeled with '+' and '*,' respectively. (**B**) Photoluminescence excitation maps of liquid exfoliated hBN for $P_1$ through $P_6$ where zero-phonon line and first phonon side band (*) are indicated with broken line boxes. The energy difference between the zero-phonon lines and the one-phonon sidebands corresponds to the dominant lattice phonon in hBN. (**C**) Scatter plot of excitation energy as a function of emission energy extracted from excitation profiles with plot of emission energy of zero phonon lines extracted from photoluminescence spectra. The broken lines indicate zero-phonon line, one-phonon sideband, and two-phonon sideband with a separation by 1370 cm$^{-1}$ (= 170 meV) corresponding to in-plane lattice phonon. Photoluminescence lifetimes (**D**) from liquid exfoliated hBN span a wide range between 1

and 6 ns, longer than that of monolayer CVD hBN, suggesting that defects in liquid exfoliated hBN are randomly distributed in depth and distance from the external surface of the crystal. Consistent with this is the observation that monolayer CVD (orange) demonstrates defects with uniformly shorter lifetimes.

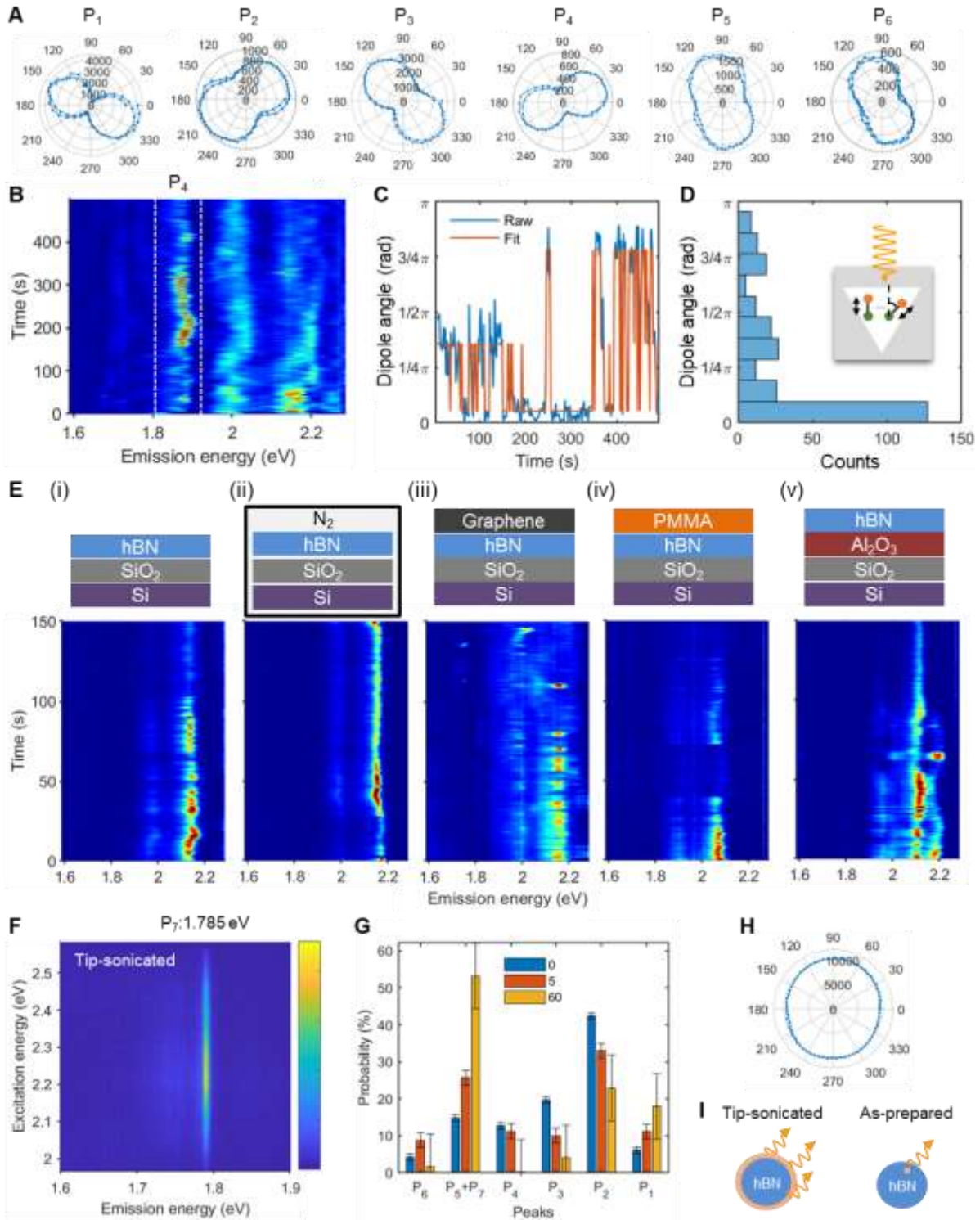

**Figure 5: Monitoring the electric dipoles of hBN vacancy centers.** Angular dependent photoluminescence intensity (**A**) of liquid exfoliated hBN, showing two-fold symmetry for $P_1$ through $P_6$ due to symmetry broken charge distribution. Time-resolved polarization measurement produces spectral trajectory (**B**) of multilayer CVD hBN around 1.88 eV ($P_4$). This type of samples exhibits blinking in intensity and spectral fluctuation for other

assignments. Extracted time traces of linear polarization angle of photoluminescence (**C**) shows typically two or three discrete states as represented in the converted histogram (**D**) from the trace (**C**), suggesting that the excitation dipole in hBN defects stochastically rotate as depicted in the inset of (**D**). Photoluminescence time trace (**E**) of multilayer hBN grown with chemical vapor deposition (CVD) with (i) no capping layer, (ii) $N_2$ purge, (iii) graphene and (iv) PMMA on top of hBN, (v) $Al_2O_3$ with atomic layer deposition underneath hBN on $SiO_2$/Si substrates. Although the $N_2$ purge and the cover layer protect the system from adsorption of additional oxygen and moisture on hBN and the $Al_2O_3$ layer insulates from charge impurities in the $SiO_2$ substrate, time traces show invariant blinking and energy fluctuations in all samples, suggesting an intrinsic mechanism. Photoluminescence excitation map (**F**) of liquid exfoliated hBN processed with tip-sonication exhibiting emission at 1.785 eV labeled as $P_7$ showing constant emission energy. Comparison of peak abundance (**G**) with tip-sonication times from 0 to 60 min, showing the evolution of $P_7$. Angular dependent photoluminescence intensity (**H**) showing that $P_7$ exhibits isotropic polarization of the electric dipole in contrast to the two-fold symmetry in the other assigned defects (**A**). Schematic of hBN crystals (**I**) hosting multiple defects along the edge in tip-sonicated hBN and a single defect in as-prepared hBN.

# Supplementary Material for "Observation and Spectral Assignment of a Family of Hexagonal Boron Nitride Lattice Defects"

*Daichi Kozawa[1,†], Ananth Govind Rajan[1,†], Sylvia Xin Li[1], Takeo Ichihara[1], Volodymyr B. Koman[1], Yuwen Zeng[1], Matthias Kuehne[1], Satish Kumar Iyemperumal[1], Kevin S. Silmore[1], Dorsa Parviz[1], Pingwei Liu[1,2], Albert Tianxiang Liu[1], Samuel Faucher, Zhe Yuan, Wenshuo Xu[3], Jamie H. Warner[3], Daniel Blankschtein[1] and, Michael S. Strano[1,*]*

[1] Department of Chemical Engineering, Massachusetts Institute of Technology, Cambridge, Massachusetts 02139, United States
[2] College of Chemical and Biological Engineering, Zhejiang University, Hangzhou, 9 Zhejiang Province, China, 310027
[3] Department of Materials, University of Oxford, Parks Road, Oxford, OX1 3PH, United Kingdom

* Corresponding Author: Michael S. Strano (E-mail: strano@mit.edu)

## Materials and Methods

### 1. Sample Preparation.

Liquid exfoliated multilayer hexagonal boron nitride (hBN) water/ethanol dispersion, chemical vapour deposition (CVD) multilayer and monolayer hBN grown on Cu foils were obtained from Graphene Supermarket. The liquid exfoliated hBN was drop-cast onto $SiO_2$/Si substrates and annealed at 850 °C for 30 min with 50 sccm of Ar flow under 1 Torr vacuum. To transfer the CVD samples onto $SiO_2$/Si substrates, we spin-coated PMMA (950 PMMA A4, MicroChem) at 2000 rpm for 2 min, baked it on a hot plate at 150 °C for 10 min to stretch wrinkles of the PMMA film, and dissolved the PMMA/hBN bilayer film in sodium persulfate Cu etchant (SPS-100, Transene) for typically 30 min. The PMMA support film was dissolved in acetone for 30 min and further removed with a thermal annealing at 300 °C for three h under an $H_2$/Ar atmosphere (10/50 sccm respectively) at 1 Torr. Au/Cr markers were deposited on $SiO_2$/Si using the photolithographic technique to indicate coordinates on substrates so that same spots can be measured. For boric acid treatments, the reaction was carried out in a quartz tube with a 1-inch diameter under low pressure at 1 Torr with 50 sccm of Ar flow for 30 min. hBN on a $SiO_2$/Si substrate was loaded on the centre of the furnace, and 0.5 g of solid boric acid powder in a W boat was placed on the edge of the heater in a furnace to keep the temperature around the W boat with boric acid ~ 300 °C while the temperature around hBN at 850 °C. For the boric acid ethicng experiment under milder conditions, first we prepared liquid-exfoliated hBN on multiple $SiO_2$/Si substrates as many as needed. Then we annealed the hBN with boric acid under the various conditions. For tip-sonication treatments, The liquid-exfoliated hBN dispersion was processed by a probe-tip sonicator (Qsonica, Q125) with a 3 mm probe at 4 W in an ice bath. The sonicated dispersion was drop-casted onto $SiO_2$/Si substrates and annealed to activate formed defects.

### 2. Optical Characterization.

The whole optical configuration is shown in **Fig. S1**. For confocal PL mapping, a continuous-wave 532 nm laser (Edmund, 35-072) was used on confocal configuration coupled with single photon avalanche diodes (PicoQuant, MPD), while the sample was scanned on piezo-electric stage. The excitation power for photoluminescence measurements was typically 150 µW. All the optical measurements were conducted at room temperature and ambient condition unless mentioned otherwise. For hyperspectral PL mapping, we obtained PL spectra scanning samples on a piezo-electric stage by typically 25×25 pixels in 25×25 µm area with excitation by the continuous-wave 532 nm laser and collection of photoluminescence signal by a spectrometer. The one-micrometre separation between pixels was enough to avoid overlapped excitation of identical defects with diffraction limited laser spot. Time-resolved photoluminescence decay measurements were conducted using a time-correlated single photon counting module (PicoQuant, HydraHarp 400) with the avalanche diode. Supercontinuum light source was used for excitation at 510 nm with 78 MHz repetition rate and 70 ps pulse width. The photon antibunching measurements were conducted in a Hanbury Brown and Twiss interferometer using the time-correlated single photon counting module connected with a 50:50 split multimode optical fibre, and lead to two Si-based single photon avalanche photo diodes (PDM, PicoQuant). The obtained $g^2(\tau)$ curve was normalized and fitted using a simple two-level model:

$$g^2(\tau) = 1 - a\exp\left(\frac{|\tau|}{\tau_{\text{PL}}}\right) \qquad (S1)$$

where $\tau$ is time delay time, $\tau_{\text{PL}}$ is PL lifetime, and $a$ is a fitting coefficient. The defect centres in hBN were excited with the continuous-wave 532 nm laser. For photoluminescence excitation measurements, supercontinuum laser light (NKT, SuperK EXW-12) was filtered by a tunable laser filter (Photon etc, LLTF) to sweep excitation wavelength every 1 nm. Long pass filters with cut off wavelength either 550 through 750 nm were used to cut Rayleigh scattering. For time-resolved polarization measurement, the angle of linear polarization of light was modulated rotating a half-wave plate, and photoluminescence spectra were continuously collected typically every 250 ms. Raman spectroscopy (Horiba Jobin Yvon LabRAM HR 800) was performed using continuous-wave 532 nm laser.

*3. Analysis of Photoluminescence Spectra.*

To analyse the photoluminescence spectra, the background was subtracted from the spectra, maintaining features of zero phonon lines and phonon side bands. Our peak finding algorithm allows to recognize a PL spectrum as an individual ZPL if the peak intensity is larger than a half of maximum intensity because the intensity of phonon side bands is typically lower than the half. The spectrum including multiple ZPLs was then deconvoluted into individual ZPLs using multiple Lorentzian function (Figure S###). The extracted emission energies of ZPLs were displayed in a histogram, where the data were filtered with linewidth range of 6-20 and 10-100 meV for liquid exfoliated and CVD hBN, respectively, to avoid counting noises as ZPLs. In total, 447, 492, and 313 photoluminescence spectra were collected for liquid-exfoliated, CVD multilayer, and CVD monolayer hBN, and 264, 434, and 189 ZPLs were extracted from them. The 20 meV-bin width of the histogram shown in Fig. 1F-H was determined by the upper bound of emission linewidth of liquid-exfoliated hBN. Peak position, intensity and linewidth of spectra were extracted fitting to spectra by the fitting. The series of obtained emission energy histograms were reconstructed as a histogram, where a cluster analysis was carried out with a peak finding algorithm and fitting with multiple Gaussian function to the histogram to determine the assignment $P_1$ through $P_6$.

*4. Analysis Methods of Time Traces of Photoluminescence Spectra.*

Photoluminescence intensity time traces consist of integral intensity of a zero-phonon line of CVD multilayer hBN at each time (**Fig. 4B**). The time traces were fitted with $I(\theta, \varphi) = I_0 \cos^2(\theta + \varphi)$ at each period, where $I$ is measured intensity, $\theta$ is angle of linear polarization of laser, $I_0$ is maximum intensity of the dipole, and $\varphi$ is phase shift $\varphi$ as the polarization angle of electric dipole. We define phase shift $\varphi = 0$ as the most abundant phase in an intensity time trace. This fitting produced the time traces of dipole angle (**Fig. 4C**). Since these time traces show discrete change with various time, we performed cluster analysis applied to the time traces to extract states, using typical machine learning algorithm implemented in MATLAB. The Silhouette method determined the most plausible number of clusters. The Gaussian Mixture Model method deconvoluted the time traces into the determined number of clusters as states. The determined states were superposed to the measured times traces as shown in orange lines of **Fig. 4C**. We define dwell time as the time that a state stays the same state and **Fig. S 37**. The distribution of dwell time shows single exponential decay and their time constants in the decay were summarized in **Table S11**.

## 5. *Characterization of hBN.*

The thickness of hBN crystals was measured with atomic force microscope (AFM) (Asylum, MFP-3D) with Si probes. The liquid exfoliated hBN with tip-sonicated hBN was characterized using X-ray photoelectron spectroscopy (XPS) (ULVAC-PHI, PHI VersaProbe II) with monochromated Al Kα source.

## 6. *Scanning Transmission Electron Microscopy.*

For the scanning transmission electron microscopy (STEM) imaging, the samples were baked overnight under high vacuum at 150 °C to reduce surface contamination. ADF-STEM was conducted using an aberration-corrected JEOL ARM300CF STEM equipped with a JEOL ETA corrector operated at an accelerating voltage of 60 or 80 kV located in the electron Physical Sciences Imaging Centre (ePSIC) at Diamond Light Source. Dwell times of 5−20 μs and a pixel size of 0.006 nm px$^{-1}$ were used for imaging with a convergence semiangle of 31.5 mrad, a beam current of 44 pA, and inner-outer acquisition angles of 49.5−198 mrad. We did not have any noticeable defect formation during our single scan imaging, which suggest that most likely, the defects are intrinsic to the samples. The ADF-STEM imaging method gave a high contrast depending on atomic mass, shown in the STEM image of the monolayer hBN (**Fig. 2C** and **Fig. S22**). Obtained STEM images were post-processed with two-dimensional fast Fourier transform filter to remove high frequency noise.

We suppressed the electron beam voltage up to 80 kV, although this is close to the energy threshold for knock-on damage in hBN lattice [1]. To evaluate the damage, we scanned the same region of monolayer CVD hBN three times by the beam at 80 kV. **Fig. S21** displays monovacancy $N=1$ and triangular $N=4$ defects in hBN monolayer. These defects are not enlarged and lattices are not created by the beam, indicating that the defects are most likely intrinsic.

## 7. *Isomer Cataloguing Framework.*

In a recent publication from our group [2], we introduced a theoretical framework, called the *Isomer Cataloguing Framework*, to predict the most-probable vacancy defects in 2D material lattices, as a function of the defect size, *N*. This framework involves combining electronic-structure density functional theory (DFT) calculations, kinetic Monte Carlo (KMC) simulations, and chemical-graph theoretic identification of various possible defect shapes, in order to determine the probabilities of occurrence of the defect shapes. Using this framework, in a previous study [2], we demonstrated remarkable agreement of the most-probable isomeric shapes of the vacancy defects in graphene, as predicted by the framework, with experimental transmission electron microscopy (TEM) images of graphene nanopores from the literature. In the Isomer Cataloguing Framework, the rates of etching ($r_{etch}$) for boron (B) and nitrogen (N) atoms in hBN are calculated using an Arrhenius expression:

$$r_{etch} = \nu \exp\left(-\frac{E_a}{k_B T}\right) \qquad (S2)$$

where $\nu = 10^{13}$ Hz is the Arrhenius prefactor [3], $E_a$ is the activation energy for the etching process, $k_B$ is the Boltzmann constant, and $T$ is the absolute temperature of the system. Note that, we have shown in our previous work [2], that altering the temperature of the system only changes the rates of formation of the nanopores, and does not alter the shapes of the most-probable isomers of the vacancy defects in the material. The activation energies for the etching of B and N atoms in hBN were calculated using a quadratic activation energy-reaction enthalpy relationship [4] and density functional theory (DFT)-based calculations of the reaction enthalpies for the etching of different types of B and N atoms in the hBN lattice, in the presence of silicon (Si) etchant atoms [2]. Indeed, our previous work [2] showed that the presence of silicon etchant atoms in essential to form the experimentally-observed triangular vacancy defects in hBN. In this regard, previous experimental work has shown that chemical vapour-deposited 2D materials could have silicon-based contaminants [5]. Moreover, recently, it has been shown that even liquid-phase exfoliated 2D materials could have silicon contaminant atoms due to them being present in the bulk powder [6]. There are mainly three types of B and N atoms in an hBN lattice with edges [7], namely, (i) a zigzag atom, which is bonded to two other atoms, each of which is bonded to two other atoms by itself; (ii) an armchair atom, which is bonded to two other atoms, one of which is bonded to two other atoms, and the second of which is bonded to two or three atoms; and (iii) a singly bonded atom. We have calculated the activation barriers of etching of these different types of B and N atoms using the etching enthalpies obtained from DFT calculations, and these values are summarized in a previous study from our group [2].

Based on the etching rates for the B and N atoms calculated using **Eq. (S1)** above, kinetic Monte Carlo (KMC) simulations [8] were carried out using a custom code written in MATLAB 2016b. The KMC simulations were initialized with a monoatom boron vacancy in the hBN lattice. Subsequently, up to 16 atoms in total (whether B or N) were removed from the hBN lattice, with the atom to be etched at each step determined using Gillespie's algorithm [9], based on the calculated etching rates. We found that the rates for etching of B and N atoms in hBN are such that, for each total number of atoms removed, there is only one *unique* nanopore, which is formed. The atomic structures of these nanopores are depicted in Fig. 2C.

### 8. DFT Calculations of the Activation Barriers for Atomic Edge Hopping and the Electronic Energy Level Gaps for Various Vacancy Defects in hBN.

Density functional theory (DFT) calculations were carried out using the Vienna Ab initio Simulation Package (VASP) [10,11], based on the projector augmented wave (PAW) formalism [12,13]. A cutoff energy of 500 eV was utilized for the plane wave basis set (see **Table S3** for convergence test), and the DFT calculations were run till the electronic energy was converged to less than $10^{-4}$ eV. PAW pseudopotentials were utilized to represent the core electrons in boron and nitrogen atoms. The Perdew-Burke-Ernzerhof (PBE) density functional [14] was utilized to represent the exchange-correlation energy of the system. Grimme's D3 method [15] was utilized to incorporate dispersion (van der Waals) interactions operative between nonbonded atoms. Reciprocal space sampling was limited to the gamma point in the *k*-space (see **Table S4** for convergence test). Gaussian smearing with a width of 0.5 eV was utilized to converge the DFT calculations. Spin polarization was considered in the calculations. All structures were geometry optimized till the maximum force on all atoms is less than 0.005 eV/Å (see **Table S5** for details regarding the geometry optimization of the primitive hBN unit cell). Climbing image-nudged elastic band (CI-NEB) calculations [16] were utilized to calculate the activation barrier for hopping of B and N atoms along the pore mouth for each nanopore formed in the KMC simulations. Three images connecting the initial and final atomic states were optimized till the total force on all atoms in the direction perpendicular to the minimum energy path was less than 0.025 eV/Å.

Electronic energies (i.e., the Kohn-Sham eigenvalues) were calculated at the gamma point, using the hybrid electronic-structure density functional, HSE06 [17-19], which is based on a screened Coulomb potential [17] and can better predict band gaps in materials [20], due to its incorporation of exact exchange interaction for electrons. For this purpose, to avoid fractional electronic occupancies in orbitals, the Gaussian smearing width was limited to 0.0258 eV. In these calculations, the electronic energy was converged to less than $10^{-8}$ eV, to obtain the electronic energy levels to high accuracy. In all our DFT calculations, a 9×9 hBN supercell was utilized (see **Fig. S19** and **Fig. S20** for a demonstration that the calculated electronic energy level gaps do not depend significantly on the choice of the 9×9 hBN supercell utilized in our calculations). Note that we chose the HSE06 functional for the hybrid DFT calculations, over other available choices, such as the PBE0 [21,22] and the B3LYP, hybrid functionals [23,24]. This is because the HSE06 functional predicts the bandgap of pristine hBN in closer agreement (3%) with the experimental data (**Fig. S12**, right panel), as compared to the PBE0 functional, which predicts the bandgap of pristine hBN within 6% of the experimental value (**Fig. S12**, middle panel). Furthermore, as opposed to the non-empirical HSE06 and PBE0 functionals, the B3LYP functional is semi-empirically parameterized, specifically for organic molecules, because of which it cannot reproduce the free electron gas limit, making it unsuitable for use in calculations of solid materials [25]. The confidence interval between the hybrid DFT and experiment still strongly supports our assignment. As the reviewer pointed out, the error in estimating energy of hBN band gap is 155 meV, which is 2.7 % of the experimental band gap. Assuming that this 2.7% error can be applied to each calculated energy, the estimated errors for P1 through P6 are 60, 57, 53, 51, 48, and 45 meV, respectively. On the other hand, the observed emission energy differences between P1 and P2, P2 and P3, …, are 100, 170, 80, 90, and 120 meV. Therefore, the observed energy difference is within the accuracy of hybrid DFT.

Finally, it is noteworthy that in our calculations of the energy gaps between the lowest unoccupied electronic levels (LUELs) and the highest occupied electronic levels (HOELs), we allowed for cross-spin electronic transitions, i.e., transitions between the spin up and spin down electronic levels. In this regard, we have verified that our assignment of defects to emission energies does not change significantly if cross-spin electronic transitions are disallowed. Indeed, when cross-spin transitions are disallowed (as opposed to what was done in the main manuscript), only the HOEL-LUEL gap for the $N$=7 vacancy defect changes from 1.94 eV, i.e., $P_4$ to 2.10 eV, i.e., $P_3$. The HOEL-LUEL gaps, and consequently the assignments, for the other vacancy defects remain unchanged.

## 9. *Reaction Analysis of Boric Acid Etching.*

To understand the transition between the various lattice vacancy, we analysed the reaction of boric acid etching to hBN lattice. The reaction kinetics of the monovacancy density $C_1$ is described with the first- and zero-order reactions:

$$\frac{C_1(t)}{dt} = -k_b C_1(t) + k_a C_0 \quad (S3)$$

where $t$ is time, $k_a$ and $k_b$ are the reaction rate constants to etch pure hBN lattice and triangular defects, and $C_0$ is the atom density of hBN lattice. The analytical solution of the differential equation is

$$C_1(t) = \frac{C_0 k_a + e^{k_b(C_1(0)-t)}}{k_b} \quad (S4)$$

The constant term is simplified by substituting $r_0$ for $\frac{C_0 k_a}{k_b}$:

$$C_1(t) = r_0 + \exp\frac{(k_b(C_1(0)-t))}{k_b} \quad (S5)$$

The **Fig. S29** A shows the model curve. We consider the temperature dependence of the reaction kinetics using the Arrhenius equation:

$$k_b = k_{b0} e^{-\frac{E_b}{k_B T}}, k_a = k_{a0} e^{-\frac{E_a}{k_B T}} \quad (S6)$$

where $k_{b0}$ and $k_{a0}$ are Arrhenius prefactors of triangular defect and hBN lattice, $E_b$ and $E_a$ are these activation energies, $k_B$ is the Boltzmann constant, and $T$ is the annealing temperature. Substituting them for Eq. S3:

$$\frac{C_1(t)}{dt} = -k_{b0} e^{-\frac{E_b}{k_B T}} C_1(t) + k_{a0} e^{-\frac{E_a}{k_B T}} C_0 \quad (S7)$$

The analytical solution is written as

$$C_1(t) = \frac{1}{k_{b0}} \exp\left(-\exp\left(-\frac{E_a}{k_B T}\right) k_{b0} t - \frac{E_a}{k_B T}\right) \left(C_1(0) \exp\left(\frac{E_a}{k_B T}\right) k_{b0} \right.$$
$$\left. + \exp\left(\frac{E_b}{k_B T}\right)\left(-1 + \exp\left(\exp\left(-\frac{E_b}{k_B T}\right) k_{b0} t\right) k_{a0}\right)\right) \quad (S8)$$

Applying the model to the experimental defect density with the annealing time, we extracted the fitting parameters as $C_1(0) = 6.71 \times 10^{-2}$ 1/μm$^2$, $k_{b0} = 2.46 \times 10^9$ 1/min, $k_{a0} = 1.88 \times 10^{10}$ 1/min, $E_b = 2.30$ eV, and $E_a = 2.86$ eV. Therefore, we obtain $k_b = 0.125$ 1/min and $k_a = 0.00288$ 1/min.

*3-Parameter Fitting to Reaction Kinetics.* Here we introduce the third reaction rate constant, $k$ for etching rate of non-triangular defects. The series of ordinary differential equations of defect density for $N$ vacancies can be written as

$$\frac{dC_{i+1}(t)}{dt} = -k_{i+1} C_{i+1}(t) + k_i C_i(t) \; [i = 1,2,\ldots,17] \quad (S9)$$

$$\begin{cases} k_i = k_{tri} \; [i = N^2] \\ k_i = k \; [i \neq N^2] \end{cases}$$

The numerical solutions of the equations $C_{i+1}(t)$ were then converted into the defect density of the assignments. Defect density of the dark states beyond the detection range in energy, $P_{dark}$ is calculated as

$$P_{dark} = 1 - \sum_{i=1}^{6} P_i \quad (S10)$$

Solid lines of Fig. 3E and **Fig. S30** display the calculated chemical reaction model at 850 °C and 220 °C, respectively. To determine $k$, we adopted the values of $k_b$ and $k_a$ extracted from the fitting to experimental data of $N_1$ defect densities shown in Figure 3C. In other words, we fitted the model to the isothermal data at 850 °C and 220 °C with the fixed $k_b$ and $k_a$, we extracted $k$ at each temperature. It should be noted that the ramping time of half hour is not counted in our fitting, since the temperature changes dramatically from room temperature to 850 °C in 30 min, and therefore the reaction constant $k$ is also changing drastically during this time.

## 10. Random Assignments with Structural Migration using Boric Acid Etching.

First, we assign the 18 vacancy states ($N_i \in N_0, N_1, \ldots, N_{16}$, and $N_{>16}$) to the 7 groups ($P_j \in P_1, P_2, \ldots, P_6$, and $P_{dark}$) randomly (Fig. 3D). Each vacancy state $N_i$ is uniformly assigned to either of the 7 groups (*i.e.*, the probability of $N_i$ assigned to $P_j$ is the same). Among these random assignments, we only count the assignments where each group $P_j$ has at least one vacancy state $N_i$ assigned to it. To avoid overfitting, we use 3 different reaction rates $k_i$ to denote the reaction constants: $k_a$ is the initial reaction constant from pure lattice to monovacancy, $k_b$ is the reaction constant from a triangular shaped vacancy to a non-triangular shaped vacancy, and $k_c$ is the one from a non-triangular to a triangular vacancy. The root mean square of fitting error with the random assignments is shown in Fig. 3D.

## 11. 5-Parameter Fitting to Reaction Kinetics.

In our model, we have 5 first-order rate constants: $k_a, k_b, k_c, k_d, k_e$. They describe the transitions between the nanopores as per the following list:

(a) 0 to 1

(b) 1 to 2, 4 to 5, 9 to 10, and 16 to 17

(c) 2 to 3, 3 to 4, 5 to 6, 8 to 9, 10 to 11, and 15 to 16

(d) 6 to 7, 11 to 12, and 13 to 14

(e) 7 to 8, 12 to 13, and 14 to 15

Accordingly, the kinetic equations describing the time variation of the number of nanopores of each size are:

$$\frac{dN_0}{dt} = -k_a N_0$$

$$\frac{dN_1}{dt} = k_a N_0 - k_b N_1$$

$$\frac{dN_2}{dt} = k_b N_1 - k_c N_2$$

$$\frac{dN_3}{dt} = k_c N_2 - k_c N_3$$

$$\frac{dN_4}{dt} = k_c N_3 - k_b N_4$$

$$\frac{dN_5}{dt} = k_b N_4 - k_c N_5$$

$$\frac{dN_6}{dt} = k_c N_5 - k_d N_6$$

$$\frac{dN_7}{dt} = k_d N_6 - k_e N_7$$

$$\frac{dN_8}{dt} = k_e N_7 - k_c N_8$$

$$\frac{dN_9}{dt} = k_c N_8 - k_b N_9$$

$$\frac{dN_{10}}{dt} = k_b N_9 - k_c N_{10}$$

$$\frac{dN_{11}}{dt} = k_c N_{10} - k_d N_{11}$$

$$\frac{dN_{12}}{dt} = k_d N_{11} - k_e N_{12}$$

$$\frac{dN_{13}}{dt} = k_e N_{12} - k_d N_{13}$$

$$\frac{dN_{14}}{dt} = k_d N_{13} - k_e N_{14}$$

$$\frac{dN_{15}}{dt} = k_e N_{14} - k_c N_{15}$$

$$\frac{dN_{16}}{dt} = k_c N_{15} - k_b N_{16}$$

$$\frac{dN_{>16}}{dt} = k_b N_{16}$$

(S11)

We fit the 5-parameter model to the reaction kinetics (Fig. 3E and Fig. S30).

## *12. Stability of the Monoatom Boron Vacancy and the Larger Triangular Defects in hBN.*

We calculated the lifetimes of the most-probable vacancy defects in hBN, for each size of the vacancy defect, *N*, as characterized by the total number of B and N atoms removed from the hBN lattice to form the vacancy defect, with *N* ranging from *N*=1 through *N*=16. The lifetimes were calculated using the Arrhenius equation, **Eq. (S2)**, stated above, at a temperature of 500 °C, using the activation barriers for etching different types of B and N atoms in hBN, as calculated in our previous work [2]. For this purpose, we considered the type of the atom that needs to be etched from the hBN lattice to form the next larger nanopore, as tabulated in **Table S2**. For example, from an examination of **Fig. S13A**, for an *N*=4 nanopore to be converted into an *N*=5 nanopore, a zigzag nitrogen (NZZ) atom needs to be etched from the hBN lattice. We find that the monoatom boron vacancy, and all subsequent triangular pores (*N*=4, 9, 16, …) are the most-stable defects, based on their largest lifetimes amongst all the pores considered. This is because, the conversion of these most-stable defects to the next largest defect requires the removal of an NZZ atom, which incurs a high activation barrier of 2.303 eV.

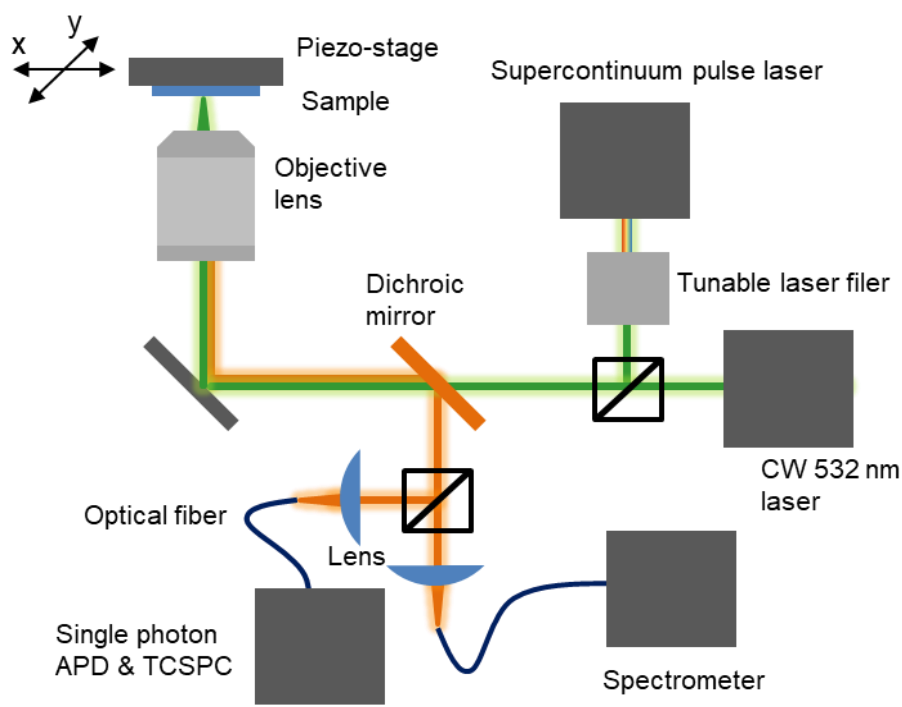

**Fig. S1.** Optical configuration for confocal and hyperspectral photoluminescence, and photoluminescence excitation spectroscopy. For excitation, continuous 532 nm solid state laser or supercontinuum pulse laser monochromated with tunable laser filter was used. For detection, single photon avalanche photodiode (APD) connected with time-correlated single photon counting (TCPSC) module or spectrometer with Si array detector was used.

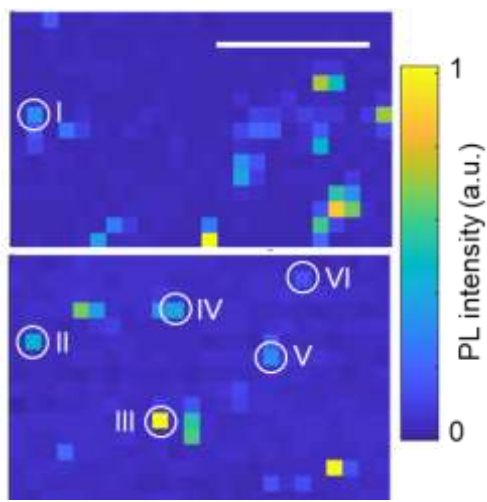

**Fig. S2.** Hyperspectral map of liquid exfoliated hBN consisting of photoluminescence spectrum scanned over 25 × 25 pixels every 1 μm (scale bar = 10 μm), where the labels of I through VI correspond to positions of emitter in the optical micrograph, atomic force micrograph, and confocal photoluminescence map shown in **Fig. 1**.

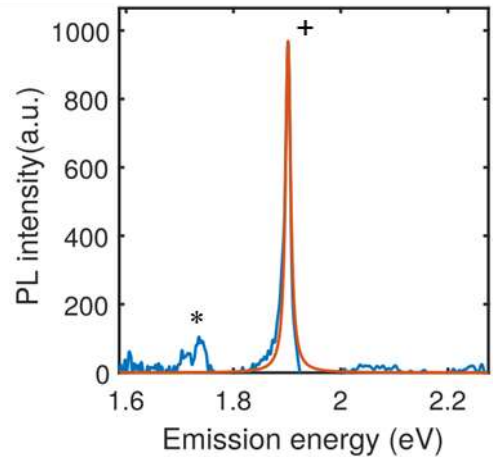

**Fig. S3. Deconvolution of ZPLs.** PL spectra of liquid exfoliated hBN to extract emission energy of ZPLs, where markers + and * indicate ZPL and phonon side band. The blue and orange curves are the measured and fit.

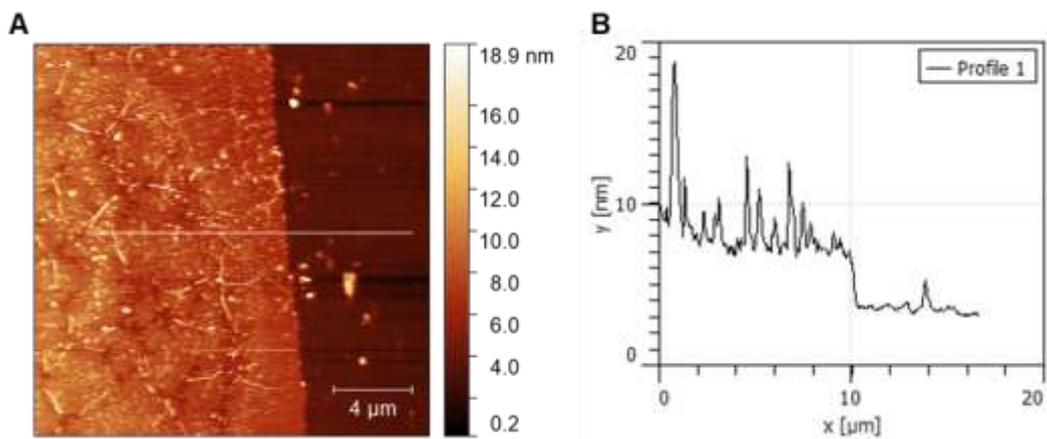

**Fig. S4.** (**A**) Atomic force micrograph and (**B**) height profiles multilayer chemical vapor deposited (CVD) hBN, where the profile shows 3 nm of thickness of hBN film corresponding to 4 layers.

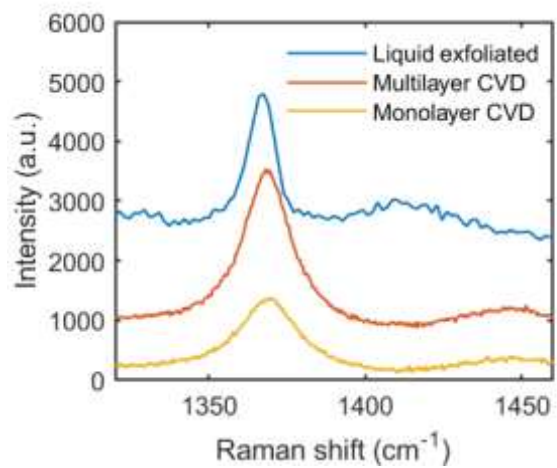

**Fig. S5.** Raman spectra of liquid exfoliated, monolayer CVD, and multilayer CVD hBN showing the phonon mode peaks at 1367.5, 1.638.2, and 1369.5 cm$^{-1}$, respectively. The spectra were vertically displaced.

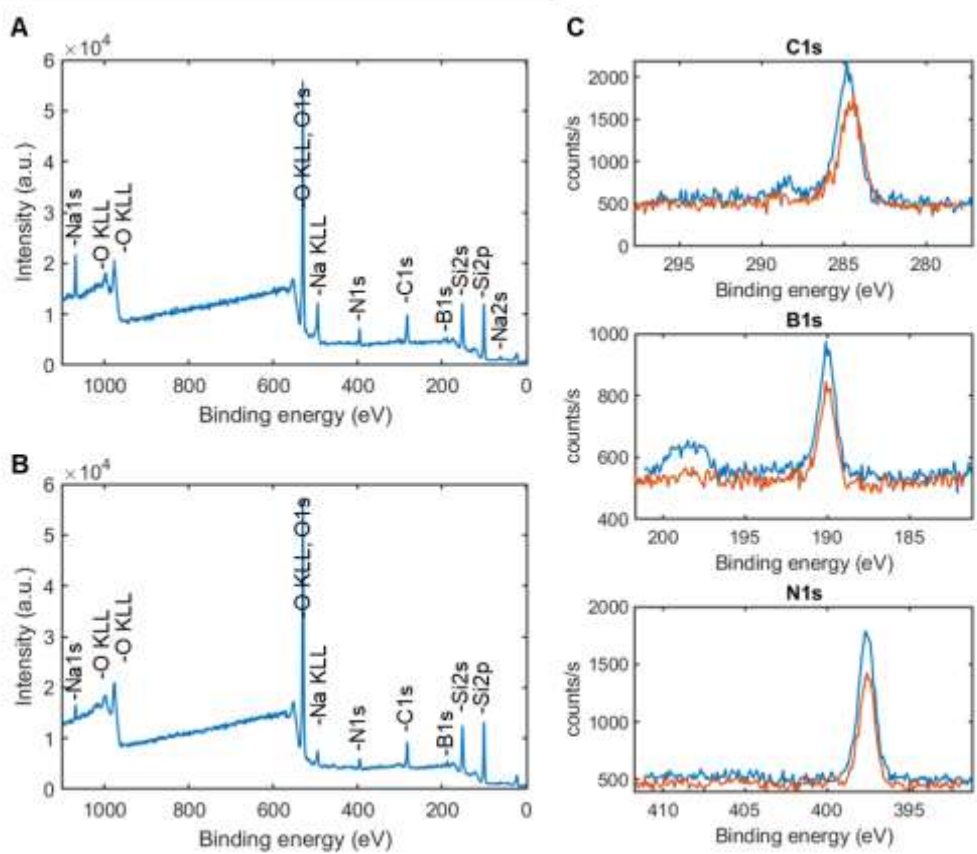

**Fig. S6.** X-ray photoelectron spectroscopy (XPS) spectra comparing between as-prepared liquid exfoliated hBN flakes (**A**) and tip-sonicated hBN flakes (**B**) on SiO$_2$/Si wafer. Both samples present residuals of Na that was attributed to the Na-containing contaminants in ethanol [26]. (**C**) Atomic peaks of carbon, boron and nitrogen don't change their shape before (blue line) and after (orange line) sonication. Atomic concentration analysis yielded 50:50 B:N ratio for both samples.

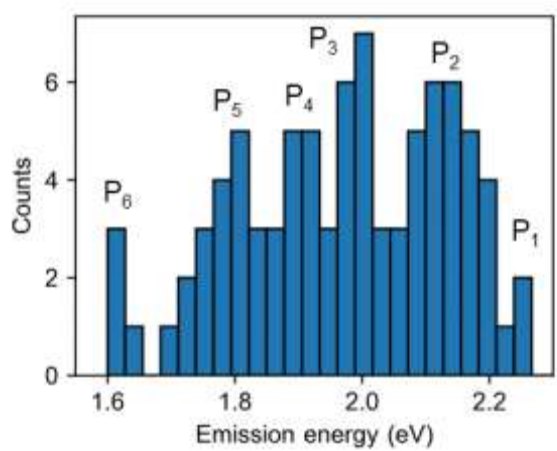

**Fig. S7.** Distribution of emission energies of 87 ZPLs observed in 19 literatures using various synthesis methods: liquid exfoliated[27-34], mechanically exfoliated[35-42], bulk crystal[43,44], CVD monolayer[33,35], CVD multilayer[45], and powder[46] hBN, showing the 6 quantized energies.

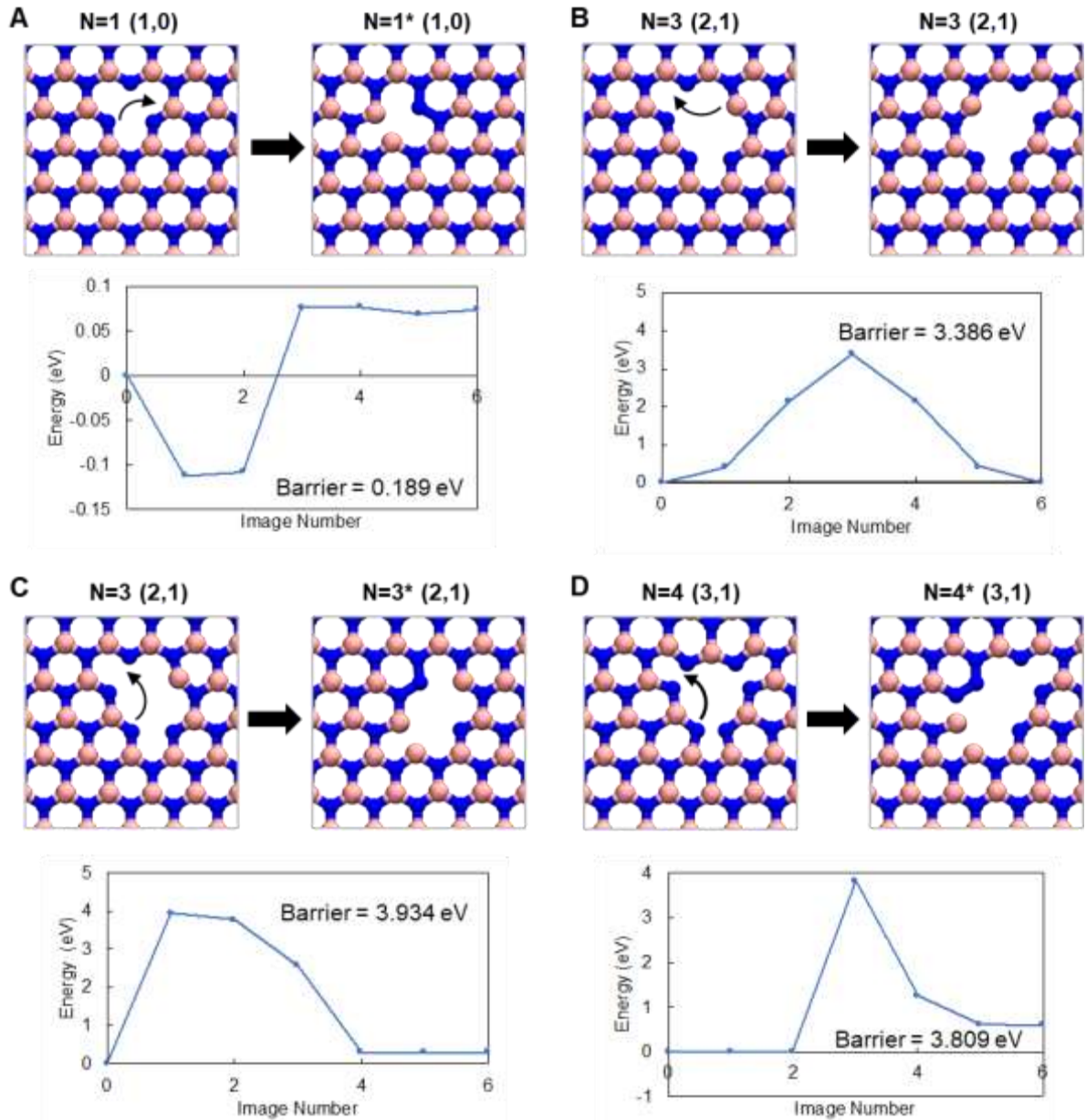

**Fig. S8. Calculated Hopping Barriers for Various Nanopore Sizes.** Atomistic models of the initial (respective left panels) and final (respective right panels) states, and the energy profile (respective bottom panels) for each depicted atomic transition from the initial state to the final state, calculated using density functional theory-based nudged elastic band calculations. (A) Nitrogen atom hopping to form a $N=1^*$ (1,0) defect from a $N=1$ (1,0) defect. (B) Boron atom hopping, leading to a 120° rotation of the $N=3$ (2,1) defect. (C) Nitrogen atom hopping to form a $N=3^*$ (2,1) defect from a $N=3$ (2,1) defect. (D) Nitrogen atom hopping to form a $N=4^*$ (3,1) defect from a $N=4$ (3,1) defect.

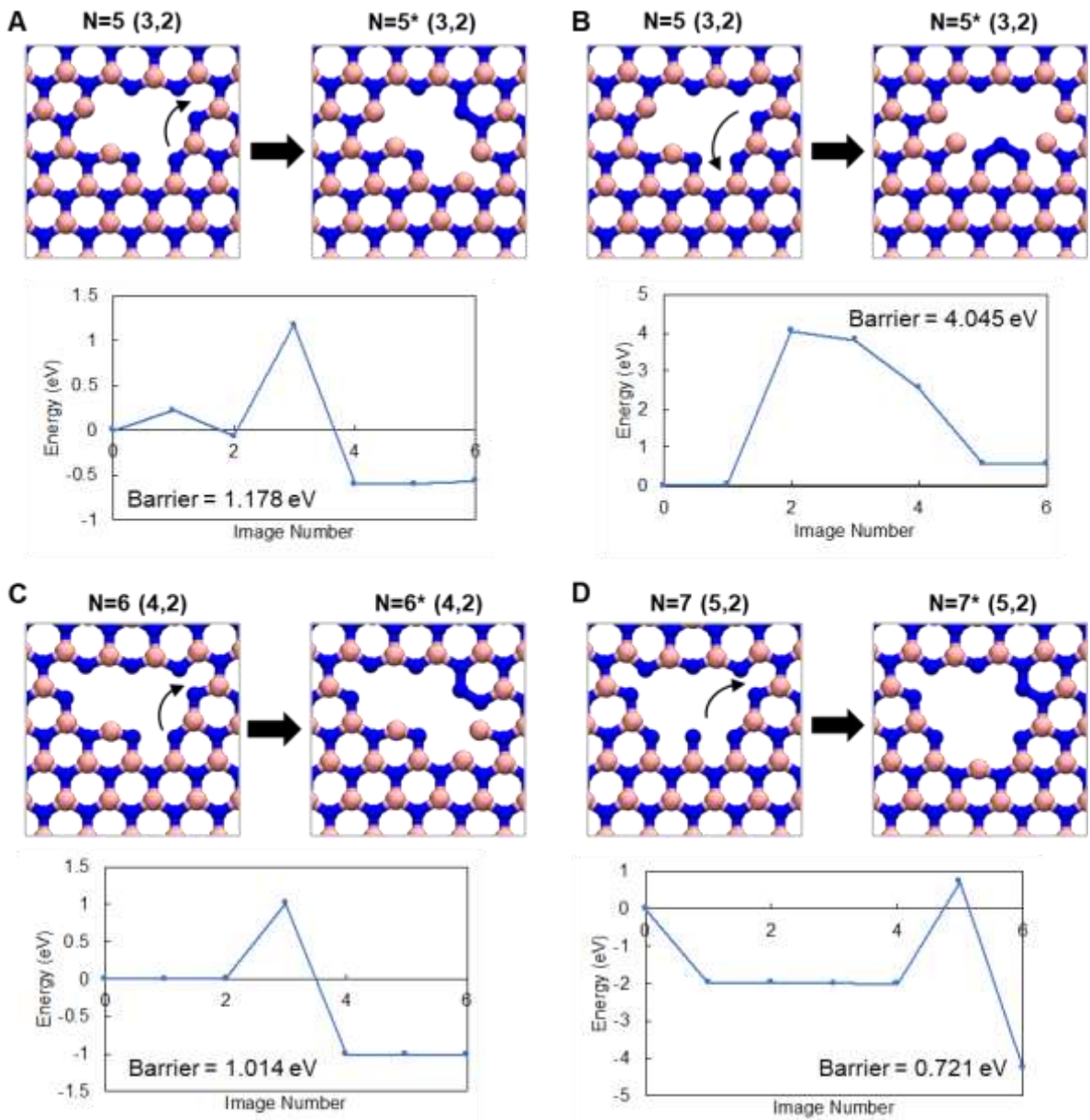

**Fig. S9**. **Calculated Hopping Barriers for Various Nanopore Sizes.** Atomistic models of the initial (respective left panels) and final (respective right panels) states, and the energy profile (respective bottom panels) for each depicted atomic transition from the initial state to the final state, calculated using density functional theory-based nudged elastic band calculations. (A) Nitrogen atom hopping to form a *N*=5* (3,2) defect from a *N*=5 (3,2) defect. (B) Nitrogen atom hopping to form a *N*=5*$^2$ (3,2) defect from a *N*=5 (3,2) defect. (C) Nitrogen atom hopping to form a *N*=6* (4,2) defect from a *N*=6 (4,2) defect. (D) Nitrogen atom hopping to form a *N*=7* (5,2) defect from a *N*=7 (5,2) defect.

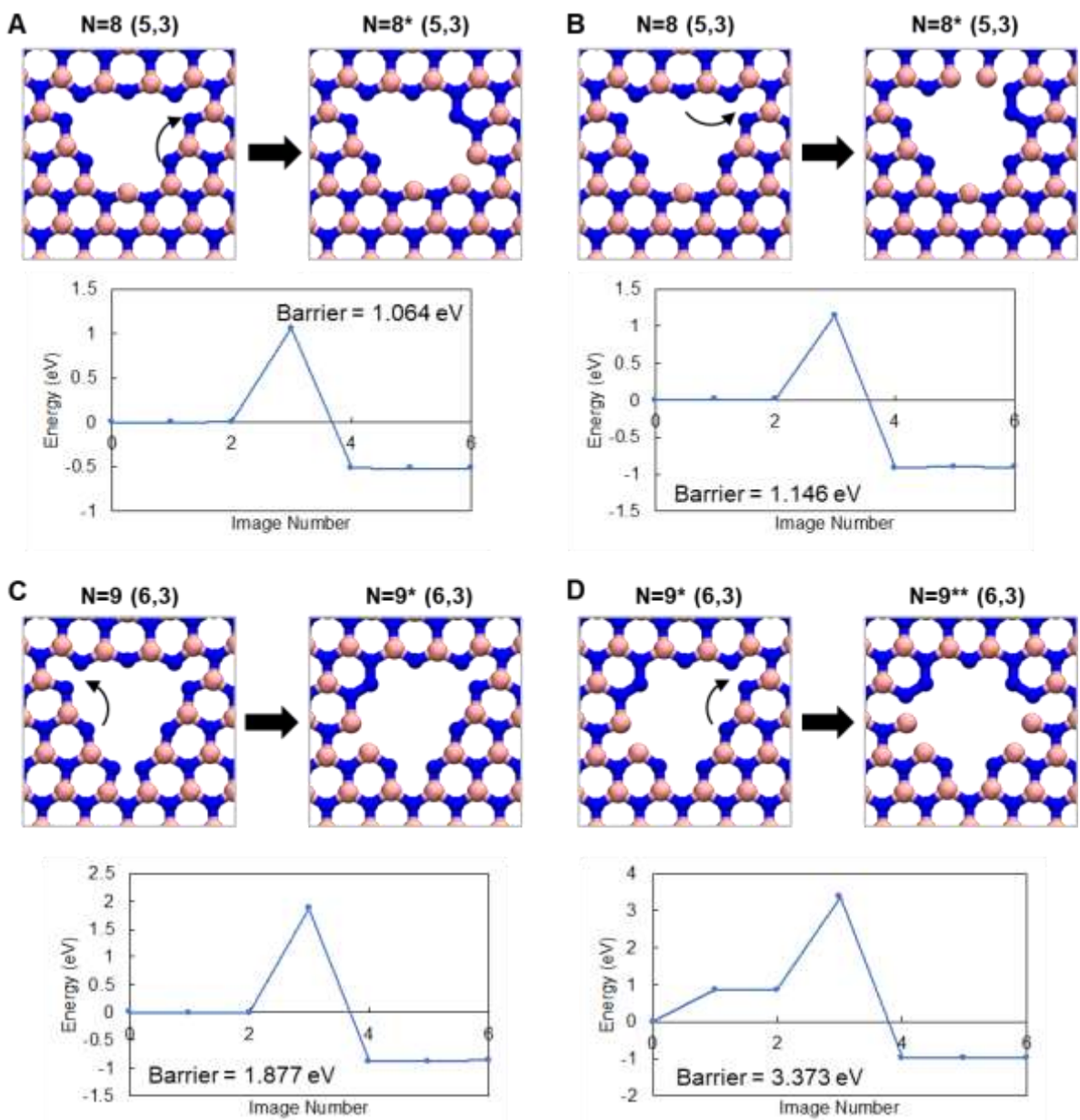

**Fig. S10. Calculated Hopping Barriers for Various Nanopore Sizes.** Atomistic models of the initial (respective left panels) and final (respective right panels) states, and the energy profile (respective bottom panels) for each depicted atomic transition from the initial state to the final state, calculated using density functional theory-based nudged elastic band calculations. (A) Nitrogen atom hopping to form a *N*=8* (5,3) defect from a *N*=8 (5,3) defect. (B) Nitrogen atom hopping to form a *N*=8*$^2$ (5,3) defect from a *N*=8 (5,3) defect. (C) Nitrogen atom hopping to form a *N*=9* (6,3) defect from a *N*=9 (6,3) defect. (D) Nitrogen atom hopping to form a *N*=9** (6,3) defect from a *N*=9* (6,3) defect.

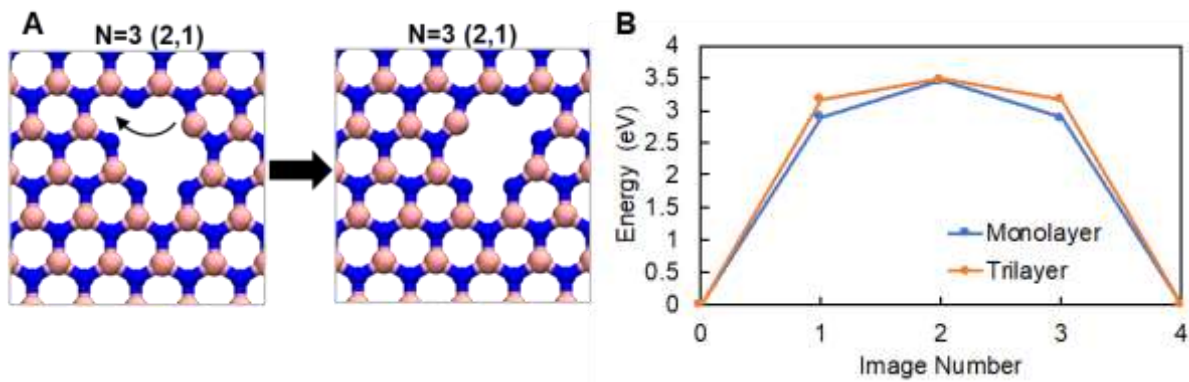

**Fig. S11. Layer Number Dependence of Calculated Barrier for Atomic Transitions.** Atomistic models of the initial (left panel) and final (right panel) states (A), and the energy profile (B) for a boron atom hopping and leading to a 120° rotation of a *N*=3 (2,1) defect in hexagonal boron nitride (hBN). The energy profiles are depicted in two cases: (i) for monolayer hBN (blue colour), wherein the defect is present in a single, isolated sheet of hBN, and (ii) for trilayer hBN (orange colour), wherein the defect is present in the middle layer, sandwiched by two hBN layers – one at the top and one at the bottom.

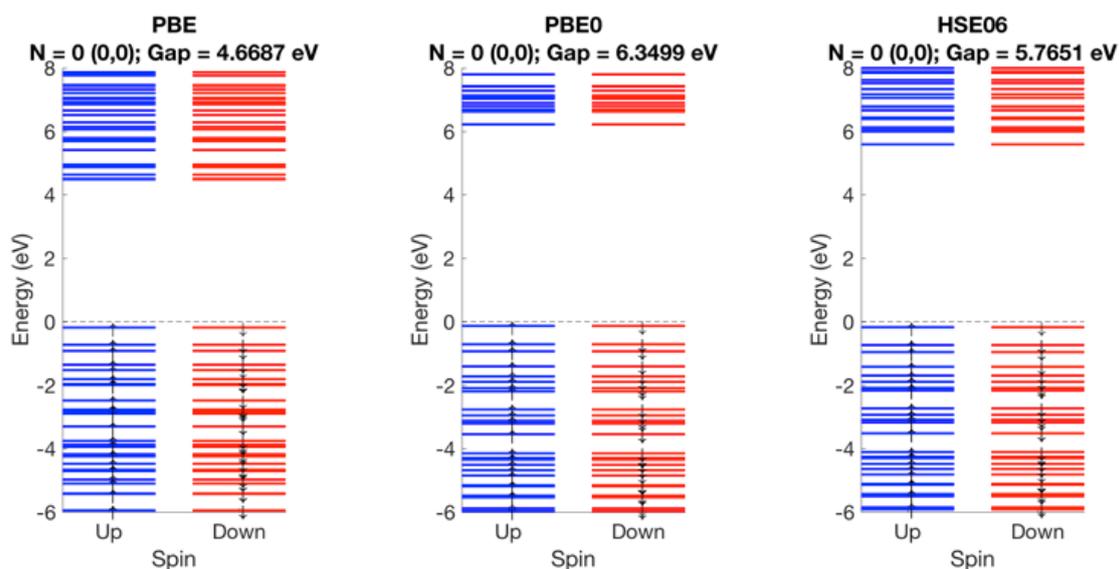

**Fig. S12.** Electronic energy-level diagram (at the $\Gamma$ point of the supercell) for pristine monolayer hexagonal boron nitride, calculated using the Perdew-Burke-Ernzerhof (PBE) generalized gradient approximation density functional (left panel), the PBE0 hybrid functional (middle panel), and the Heyd-Scuseria-Ernzerhof (HSE06) hybrid density functional (right panel), for a 9×9 supercell. Spin-up states are shown in blue colour and spin-down states are shown in red colour. Filled spin-up states are indicated with an up arrow (↑) and filled spin-down states are indicated with a down arrow (↓). The inferred energy gap is indicated above each energy diagram.

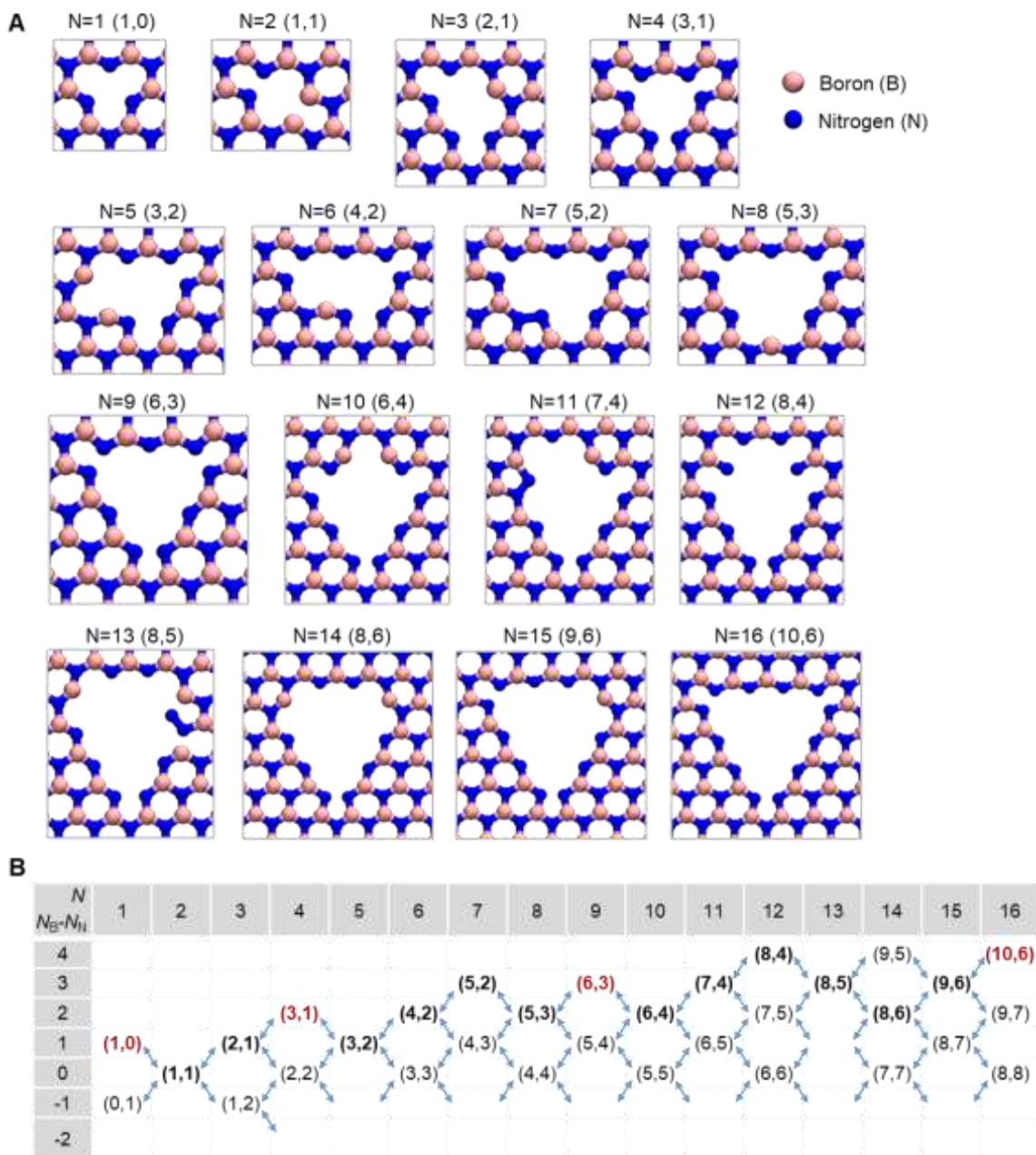

**Fig. S13.** (A) Atomic models of the most-probable isomer structures of hBN defects optimized with density functional theory for each number of vacancy atoms through $N = 1$ through 16, where $N_B$ and $N_N$ in ($N_B$, $N_N$) on the top of the models are the number of boron and nitrogen vacancy atoms. (B) Branches of possible isomers of hBN defects for $N$=1 through 16, where the most-probable isomers are indicated in bold and triangular isomers are indicated in red, showing expanding the vacancy sizes with time in this order.

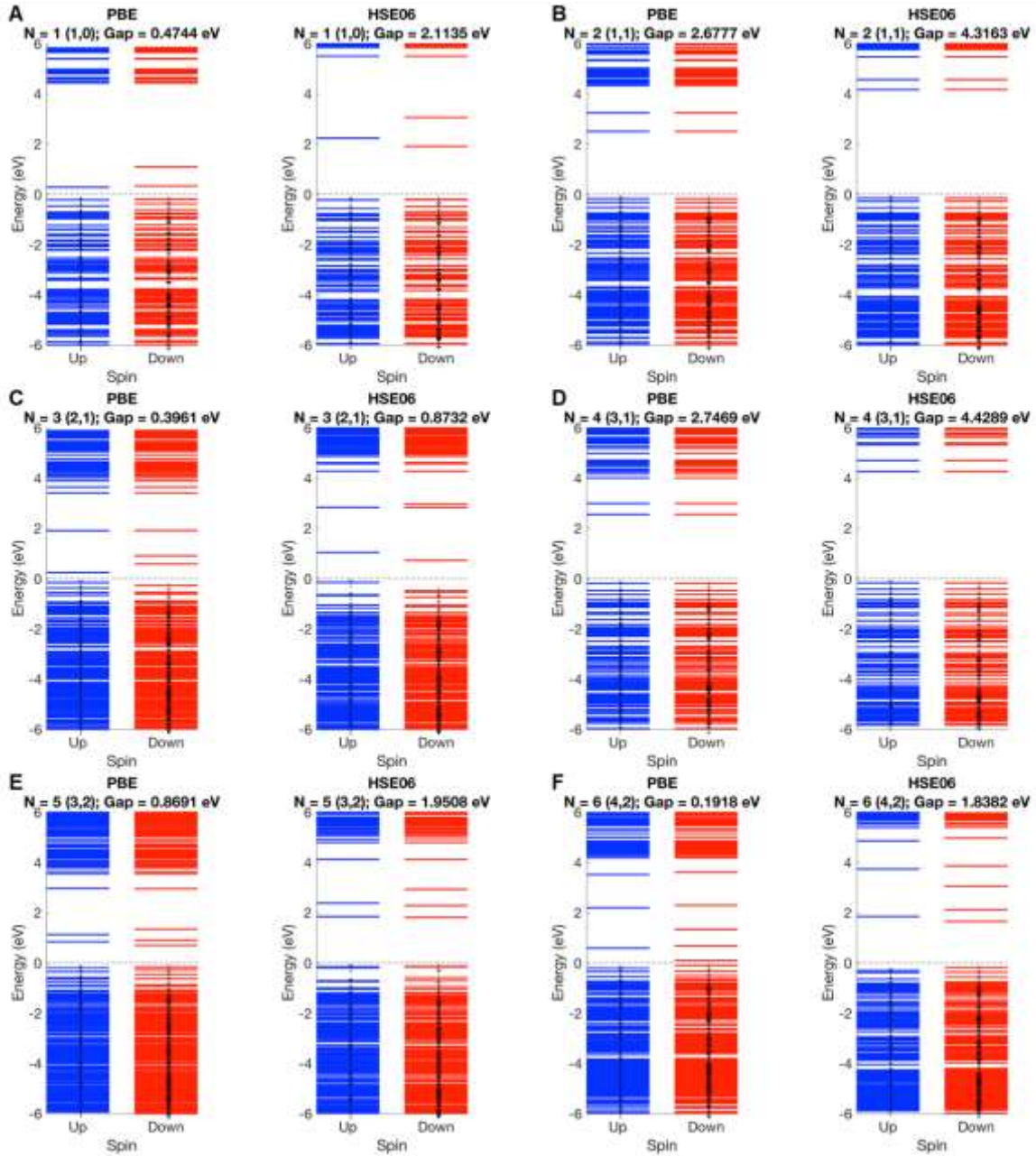

**Fig. S14.** Electronic energy-level diagram (at the $\Gamma$ point of the supercell) for the (A) $N$=1 (1,0), (B) $N$=2 (1,1), (C) $N$=3 (2,1), (D) $N$=4 (3,1), (E) $N$=5 (3,2), (F) $N$=6 (4,2) defects in monolayer hexagonal boron nitride, calculated using the Perdew-Burke-Ernzerhof (PBE) generalized gradient approximation density functional (left panel) and the Heyd-Scuseria-Ernzerhof (HSE06) hybrid density functional (right panel), for a 9×9 supercell. Spin-up states are shown in blue colour and spin-down states are shown in red colour. Filled spin-up states are indicated with an up arrow ($\uparrow$) and filled spin-down states are indicated with a down arrow ($\downarrow$). The inferred energy gap is indicated above each energy diagram.

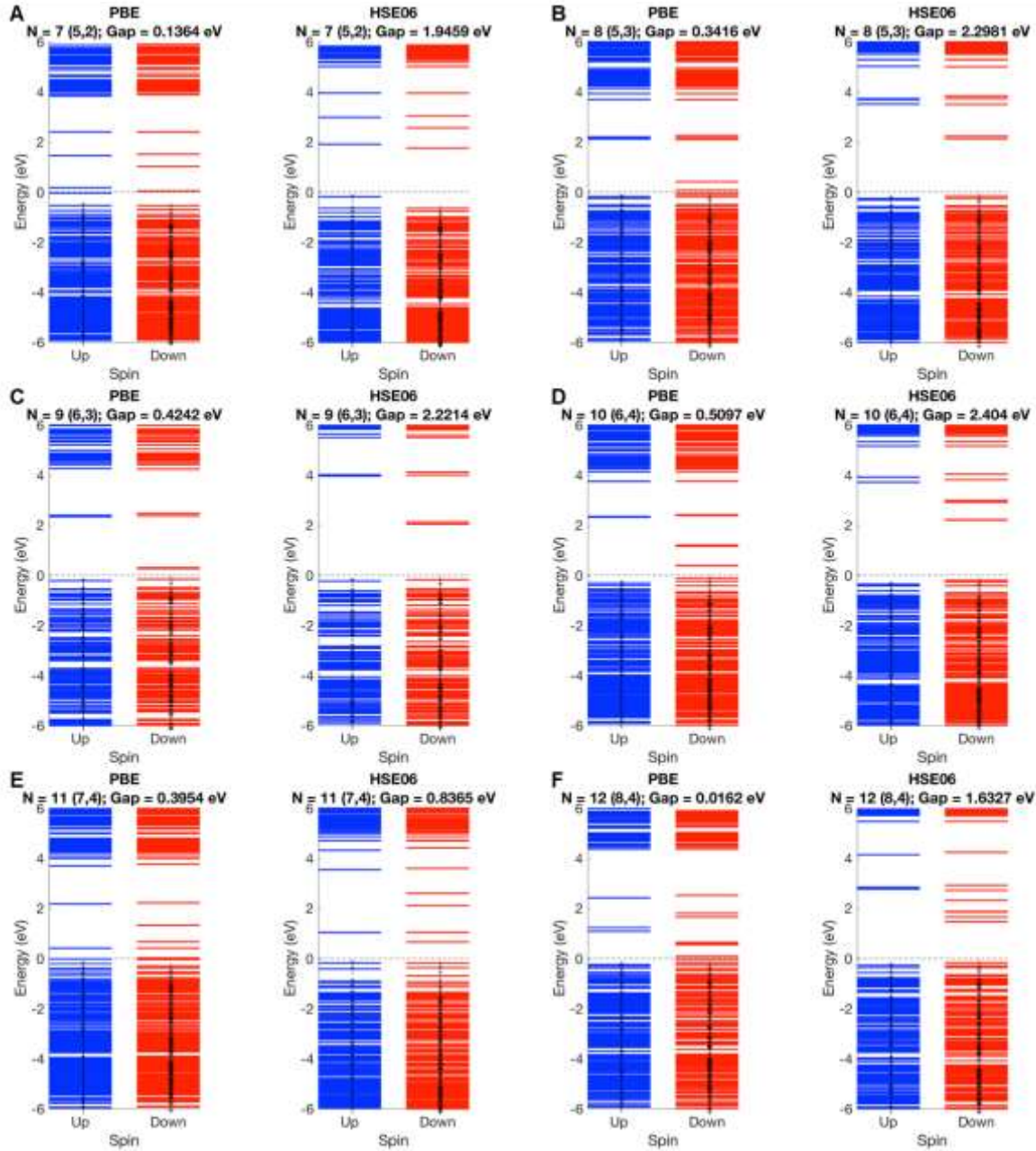

**Fig. S15.** Electronic energy-level diagram (at the $\Gamma$ point of the supercell) for the (A) $N$=7 (5,2), (B) $N$=8 (5,3), (C) $N$=9 (6,3), (D) $N$=10 (6,4), (E) $N$=11 (7,4), (F) $N$=12 (8,4) defects in monolayer hexagonal boron nitride, calculated using the Perdew-Burke-Ernzerhof (PBE) generalized gradient approximation density functional (left panel) and the Heyd-Scuseria-Ernzerhof (HSE06) hybrid density functional (right panel), for a 9×9 supercell. Spin-up states are shown in blue colour and spin-down states are shown in red colour. Filled spin-up states are indicated with an up arrow (↑) and filled spin-down states are indicated with a down arrow (↓). The inferred energy gap is indicated above each energy diagram.

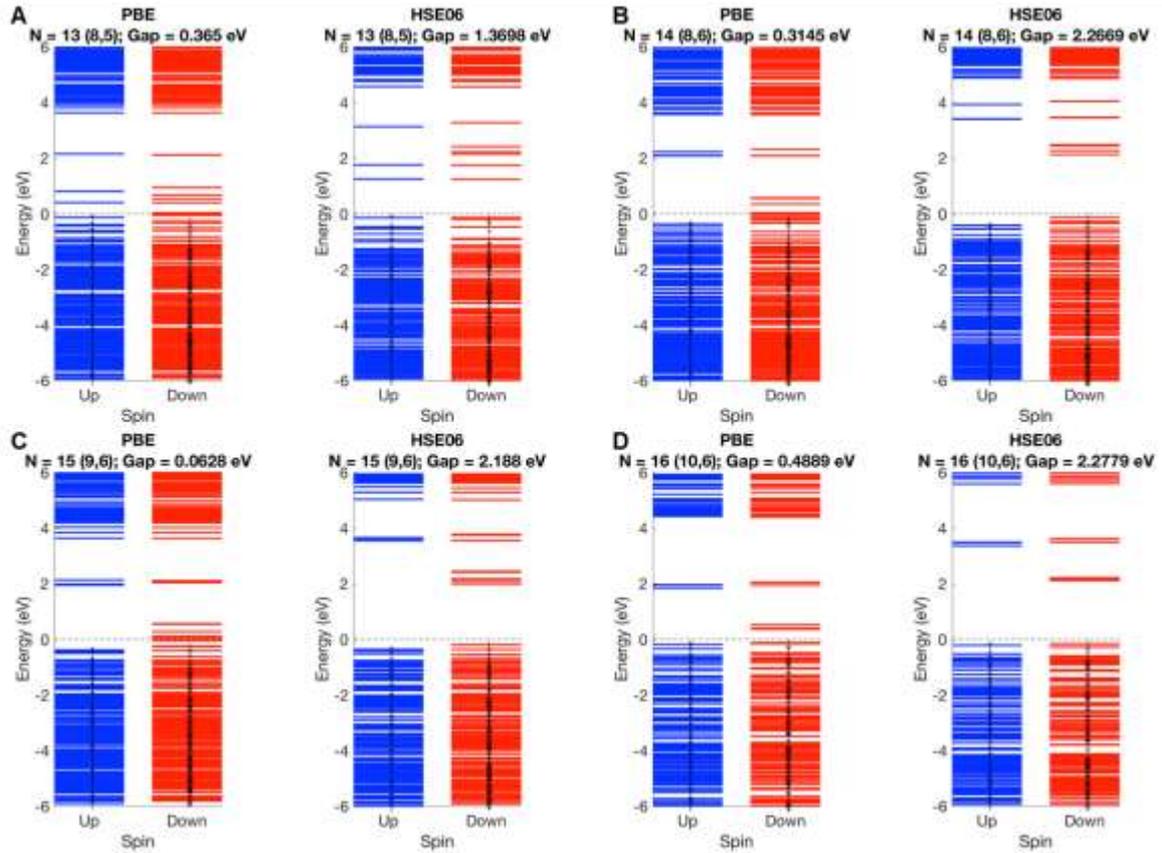

**Fig. S16.** Electronic energy-level diagram (at the $\Gamma$ point of the supercell) for the (A) $N$=13 (8,5), (B) $N$=14 (8,5), (C) $N$=15 (9,6), (D) $N$=16 (10,6) defects in monolayer hexagonal boron nitride, calculated using the Perdew-Burke-Ernzerhof (PBE) generalized gradient approximation density functional (left panel) and the Heyd-Scuseria-Ernzerhof (HSE06) hybrid density functional (right panel), for a 9×9 supercell. Spin-up states are shown in blue colour and spin-down states are shown in red colour. Filled spin-up states are indicated with an up arrow (↑) and filled spin-down states are indicated with a down arrow (↓). The inferred energy gap is indicated above each energy diagram.

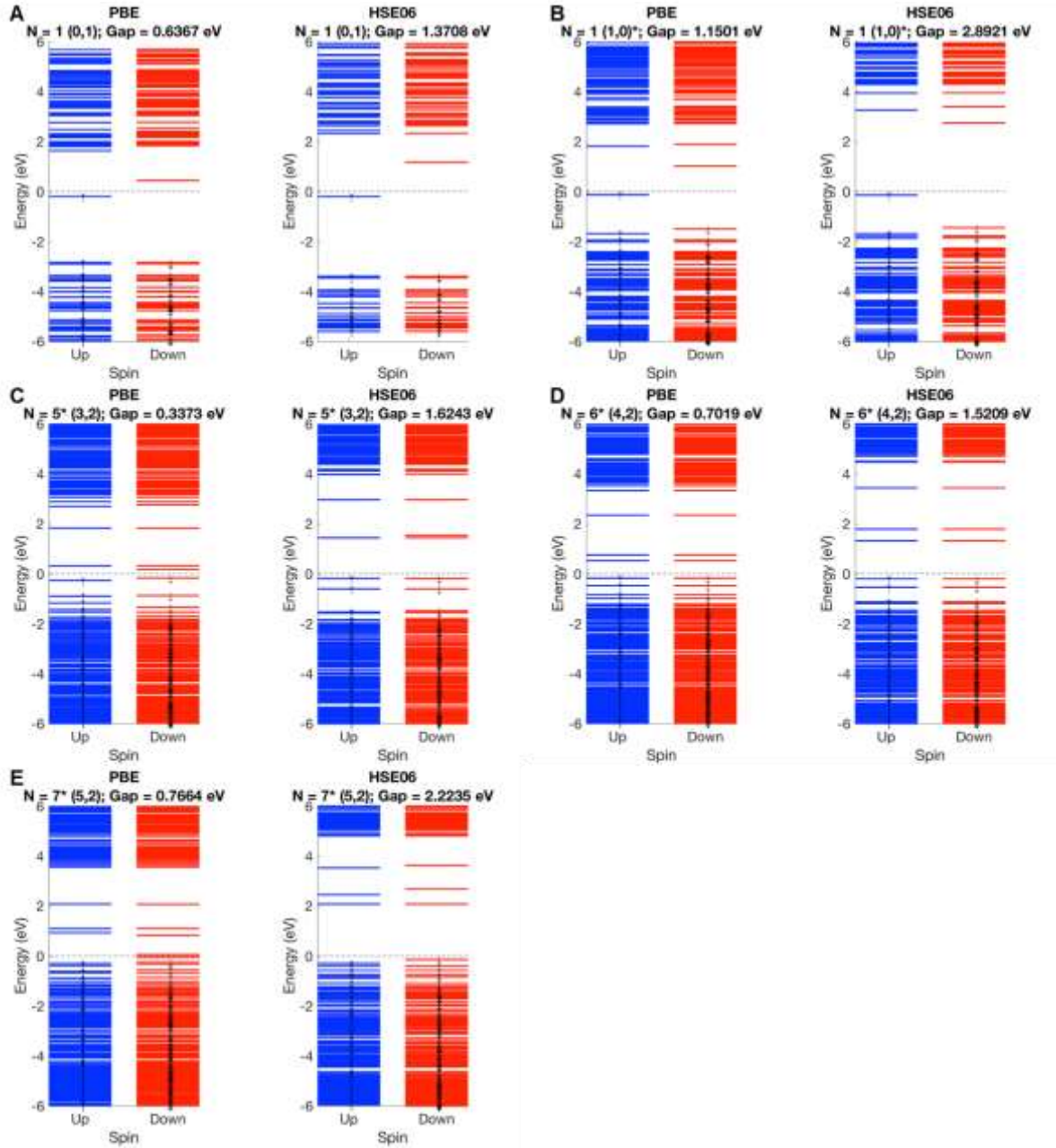

**Fig. S17.** Electronic energy-level diagram (at the Γ point of the supercell) for the (A) *N*=1 (0,1), (B) *N*=1* (1,0), (C) *N*=5* (3,2), (D) *N*=6* (4,2), (E) *N*=7* (5,2) defects in monolayer hexagonal boron nitride, calculated using the Perdew-Burke-Ernzerhof (PBE) generalized gradient approximation density functional (left panel) and the Heyd-Scuseria-Ernzerhof (HSE06) hybrid density functional (right panel), for a 9×9 supercell. Spin-up states are shown in blue colour and spin-down states are shown in red colour. Filled spin-up states are indicated with an up arrow (↑) and filled spin-down states are indicated with a down arrow (↓). The inferred energy gap is indicated above each energy diagram.

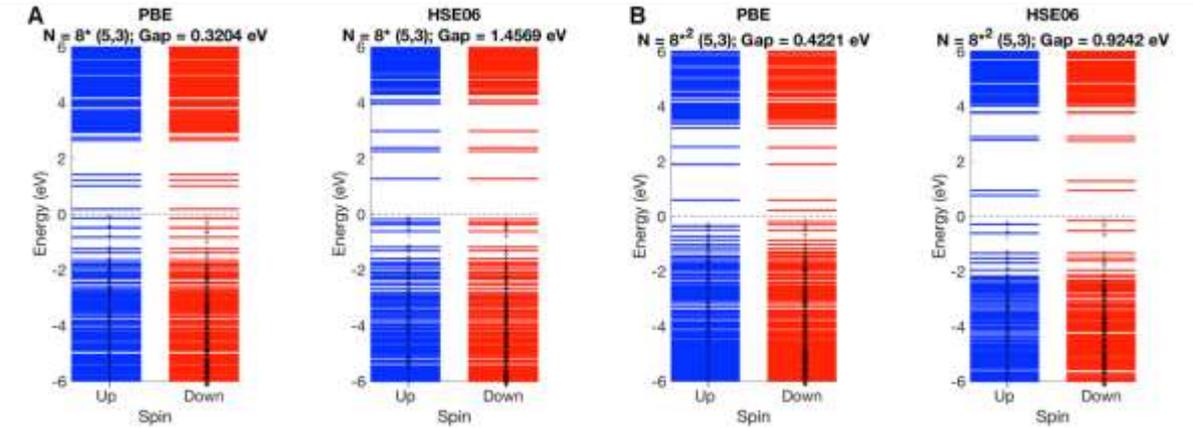

**Fig. S18.** Electronic energy-level diagram (at the $\Gamma$ point of the supercell) for the (A) $N=8^*$ (5,3), (B) $N=8^{*2}$ (5,3) defects in monolayer hexagonal boron nitride, calculated using the Perdew-Burke-Ernzerhof (PBE) generalized gradient approximation density functional (left panel) and the Heyd-Scuseria-Ernzerhof (HSE06) hybrid density functional (right panel), for a 9×9 supercell. Spin-up states are shown in blue colour and spin-down states are shown in red colour. Filled spin-up states are indicated with an up arrow ($\uparrow$) and filled spin-down states are indicated with a down arrow ($\downarrow$). The inferred energy gap is indicated above each energy diagram.

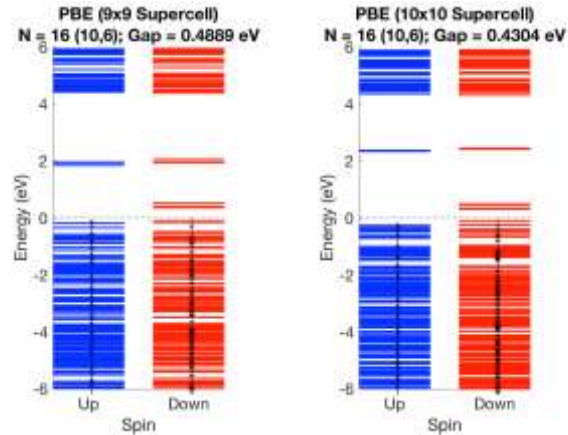

**Fig. S19. Demonstration that the Calculated Electronic Energy Gap Does Not Significantly Depend on the Supercell Size Used in the Calculations.** Electronic energy-level diagram (at the $\Gamma$ point of the supercell) of the largest defect considered in this study, i.e., the $N=16$ (10,6) defect, in monolayer hexagonal boron nitride, calculated using the Perdew-Burke-Ernzerhof (PBE) [14] generalized gradient approximation density functional, for a 9×9 supercell (left panel) and a 10×10 supercell (right panel). Spin-up states are shown in blue colour and spin-down states are shown in red colour. Filled spin-up states are indicated with an up arrow ($\uparrow$) and filled spin-down states are indicated with a down arrow ($\downarrow$). The inferred energy gap is indicated above each energy diagram. The energy levels are qualitatively similar for the two cases, and the inferred electronic energy gap differs by 0.06 eV for the two cases.

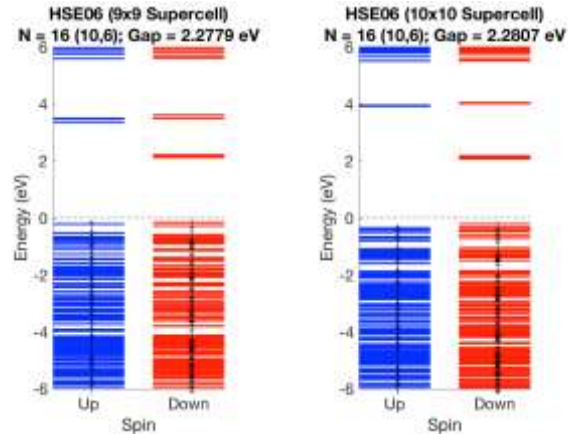

**Fig. S20. Demonstration that the Calculated Electronic Energy Gap Does Not Significantly Depend on the Supercell Size Used in the Calculations.** Electronic energy-level diagram (at the $\Gamma$ point of the supercell) of the largest defect considered in this study, i.e., the $N$=16 (10,6) defect, in monolayer hexagonal boron nitride, calculated using the Heyd-Scuseria-Ernzerof (HSE06) [17,47] hybrid exchange-correlation density functional, for a 9×9 supercell (left panel) and a 10×10 supercell (right panel). Spin-up states are shown in blue colour and spin-down states are shown in red colour. Filled spin-up states are indicated with an up arrow ($\uparrow$) and wfilled spin-down states are indicated with a down arrow ($\downarrow$). The inferred energy gap is indicated above each energy diagram. The energy levels are qualitatively similar for the two cases, and the inferred electronic energy gap differs by 0.003 eV for the two cases.

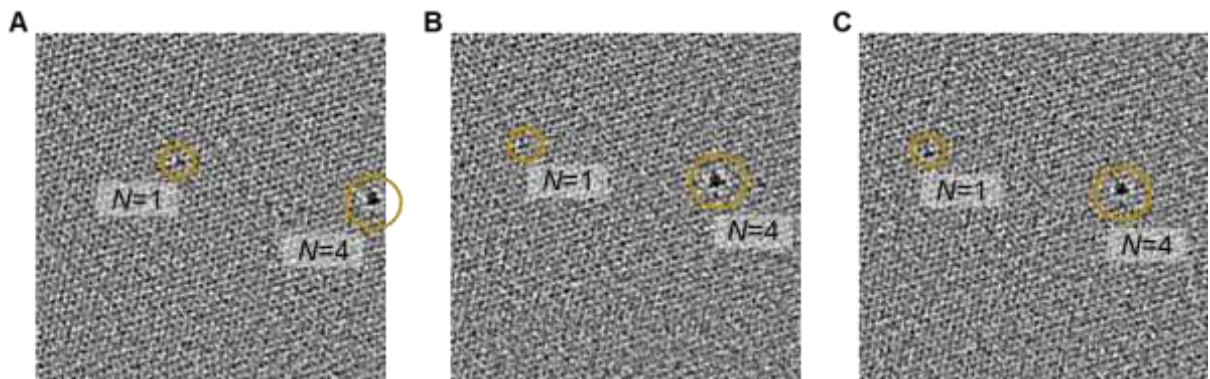

**Fig. S21. Evaluation of knock-on damage.** STEM imaging of monolayer CVD hBN lattice showing $N$=1 and $N$=4 defects.

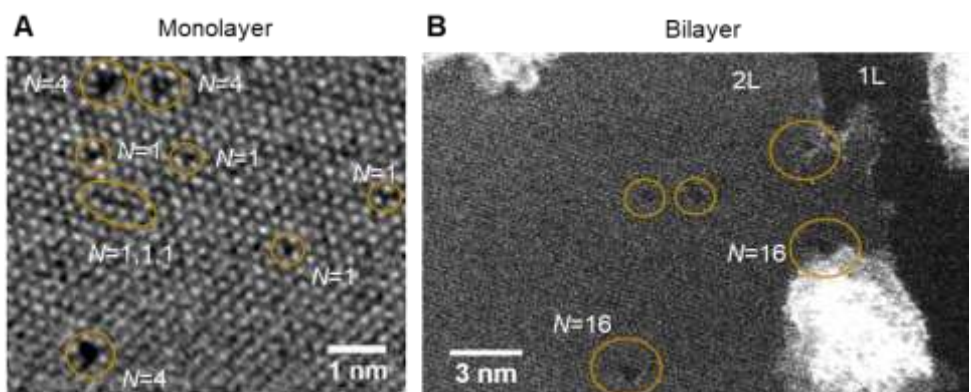

**Fig. S22.** Zoom-out view of scanning transmission electron micrographs (STEM) for (**A**) monolayer and (**B**) bilayer CVD hBN, showing $N$=1 and 4 lattice vacancies in the monolayer and $N$=16 and unassigned lattice vacancies in the bilayer. The bilayer (2L) CVD hBN also shows a monolayer (1L) region and the brighter features on the left top, right top and bottom could be carbon contaminants or polymer residues of PMMA.

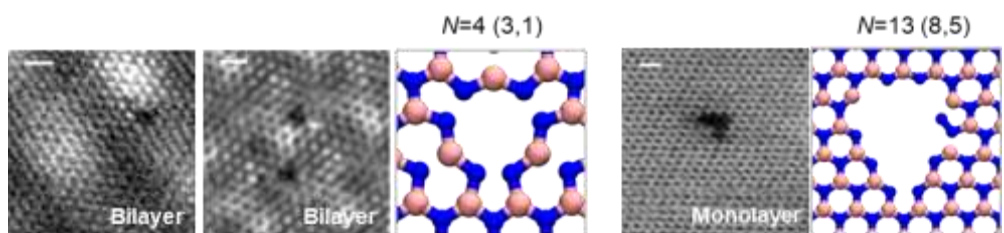

**Fig. S23.** STEM images of $N$=4 and $N$=13 observed in bilayer and monolayer CVD hBN, respectively. The scale bars are 5 Å. The corresponding atomic structures are shown on the right.

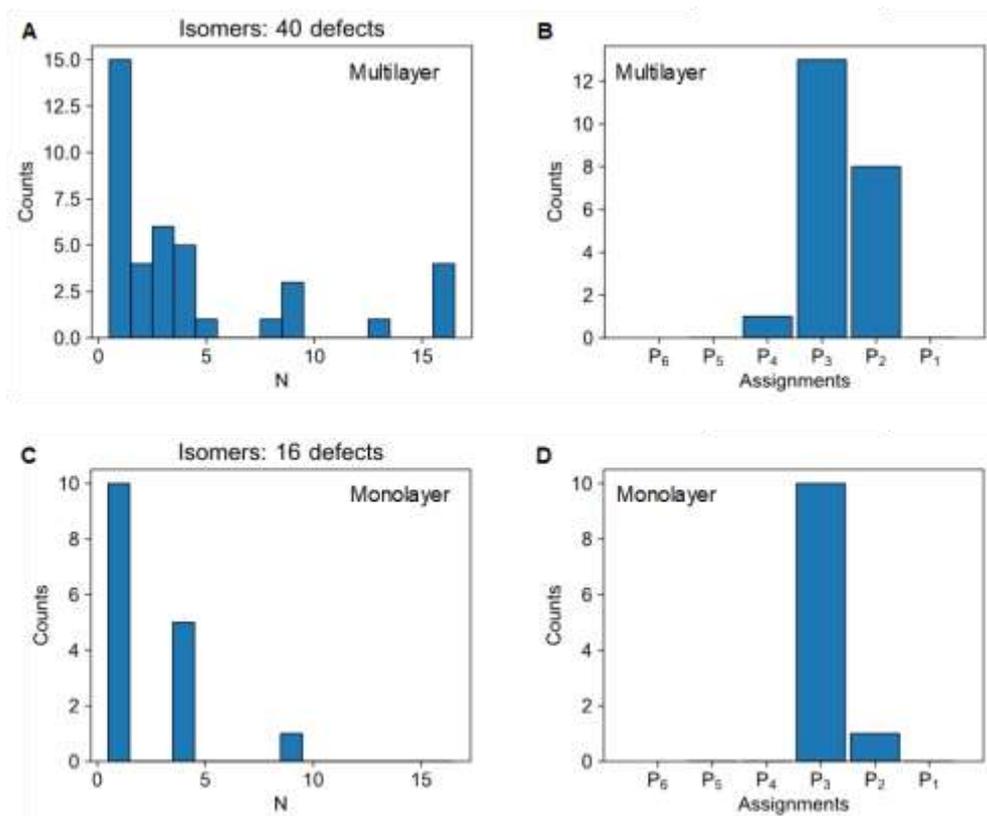

**Fig. S24.** Histograms of (**A**) the size of the most-probable isomers and (**B**) the assignment for multilayer CVD extracted from the STEM images. Histograms of (**C**) the size of the most-probable isomers and (**D**) the assignment for monolayer CVD.

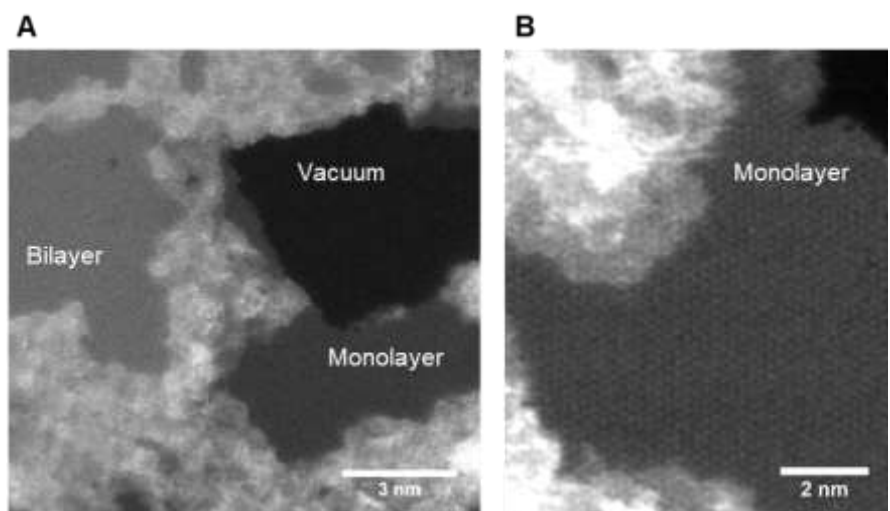

**Fig. S25.** STEM image of multilayer CVD hBN, showing vacuum, monolayer, and bilayer regions, and (B) monolayer CVD hBN without defects within the view.

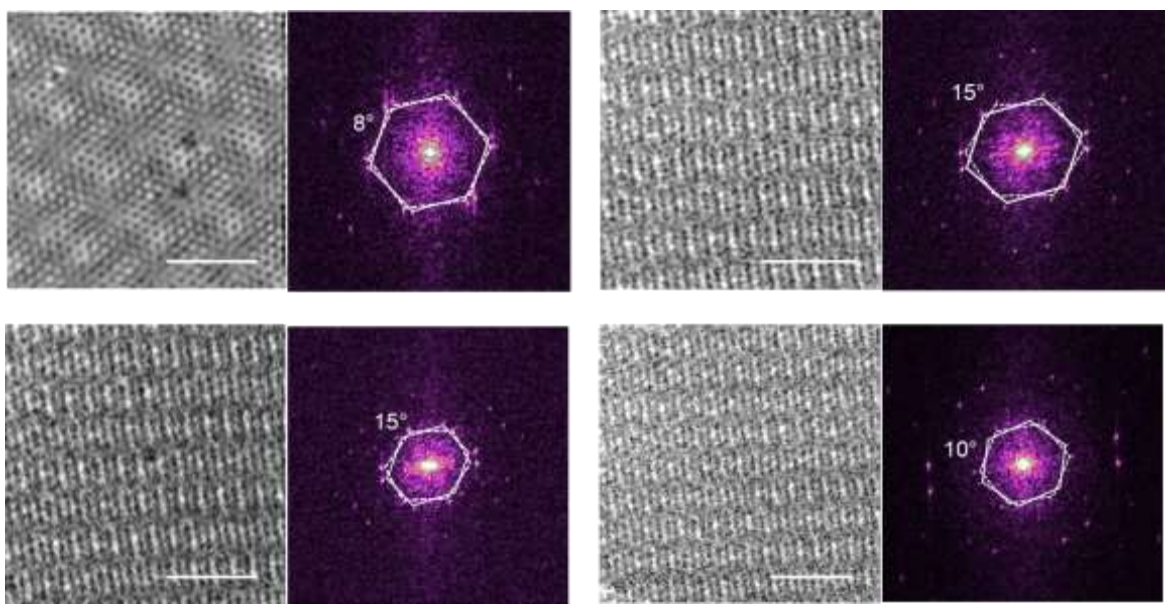

**Fig. S26.** Four different spots of (left) STEM images and (right) corresponding fast Fourier transform images of multilayer CVD hBN in the same batch. The STEM images exhibit moiré pattern due to the presence of lattice mismatching between layers, which became clear in the fast Fourier transform image showing duplicate hexagonal dots configurations. As highlighted with solid and broken lines of hexagons, the stacking angle spanned 8 through 15°. The scale bars correspond to 2 nm.

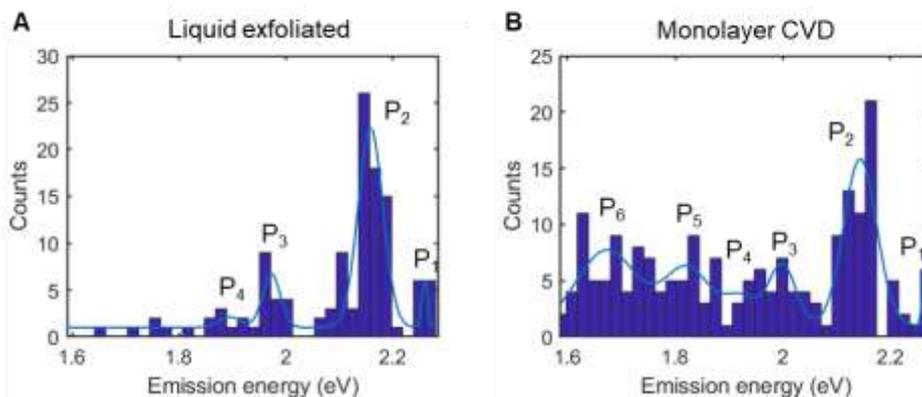

**Fig. S27.** Histogram of emission energy for (A) liquid exfoliated hBN and (B) monolayer CVD hBN treated with boric acid, fitted with multiple Gaussian function as shown by solid lines to determine the emission modes $P_1$ through $P_6$.

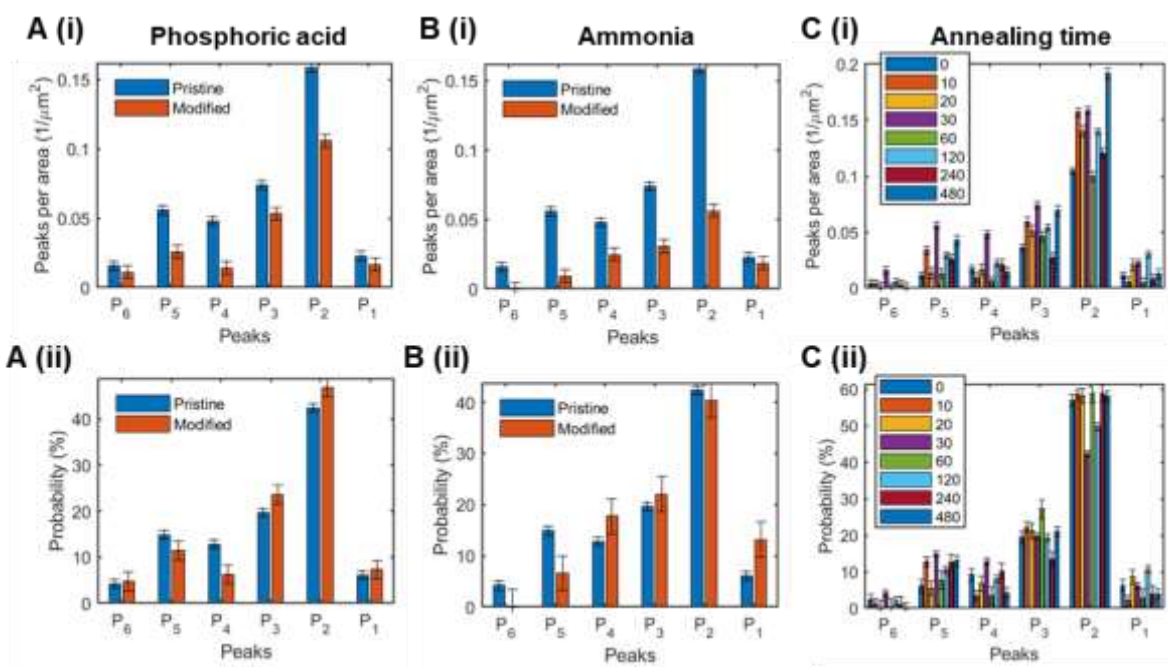

**Fig. S28.** Bar charts of frequency per unit area (i) and probability (ii) in liquid exfoliated hBN with and without (**A**) phosphoric acid, (**B**) ammonia, and (**C**) thermal annealing treatments for 0 to 480 min at 850 °C, showing only a slight change in probability.

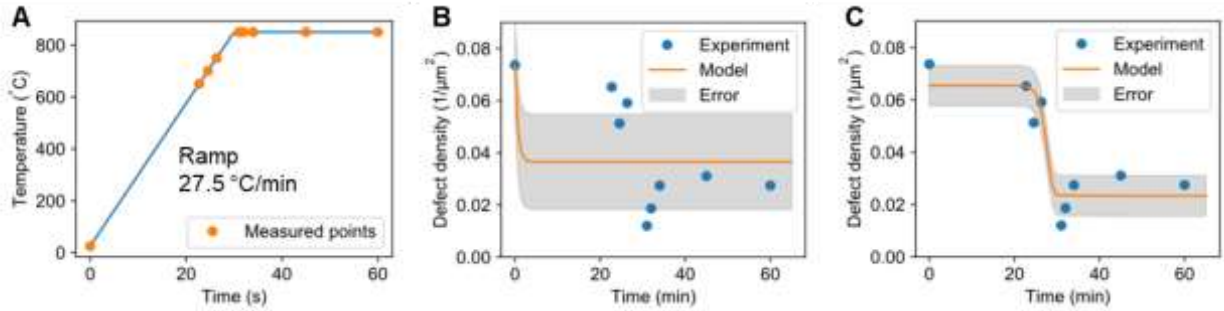

**Fig. S29.** (**A**) Set temperature and time of the boric acid etching to liquid-exfoliated hBN, gradually promoting the reaction to expands defects. Defect density of liquid exfoliated hBN with annealing time extracted from the ZPL peaks per area of the hyperspectral map. The solid lines display analytical model curves (**B**) without and (**C**) with temperature dependence of etching rates. The errors were estimated from standard deviation between the experiment and model. Raising annealing temperature from room temperature (RT) to 650, 700, 750, and 850 °C leads to increase etching rate of $P_3$. The analytical model considering temperature-dependent reaction kinetics reproduces the photoluminescence experiments to extract the defect density.

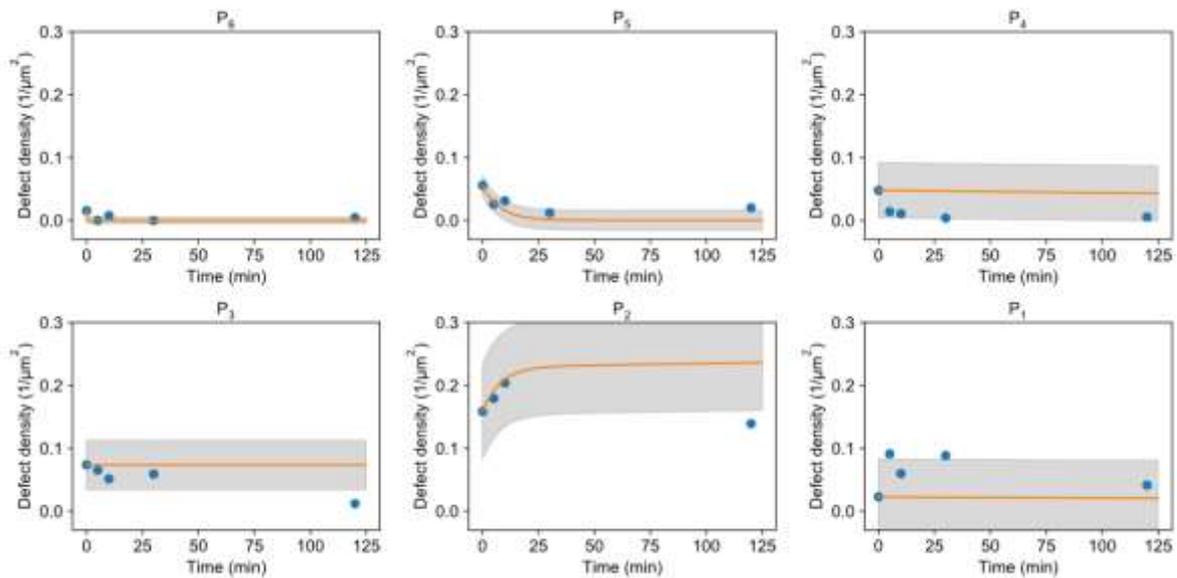

**Fig. S30.** Assigned defect density with annealing time at 220 °C. The model based on the numerical solution of rate equations describes the distribution of the assignments, showing the reaction rate constants of $k_a = 1.22 \times 10^{-19}$, $k_b = 8.83 \times 10^{-15}$, $k_c = 0.00868$, $k_d = 0.0356$ and $k_e = 1.56$ 1/min. The errors were estimated from standard deviation between the experiment and model.

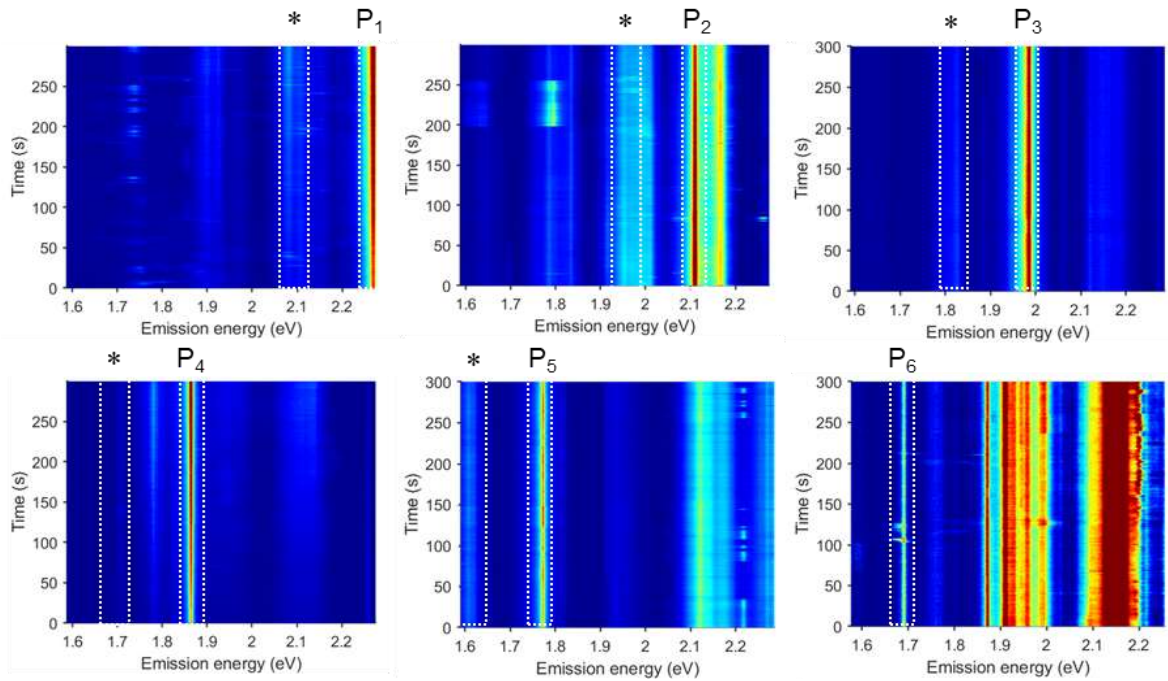

**Fig. S31.** Spectral trajectory of liquid exfoliated hBN for 300 s. The zero-phonon lines ($P_1$ through $P_6$) and one-phonon side bands are highlighted with dotted boxes, showing absence of blinking.

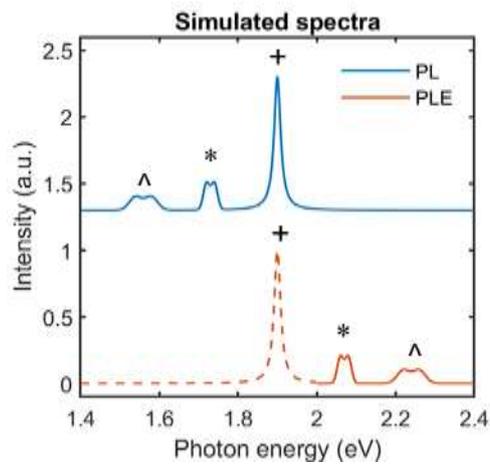

**Fig. S32.** As the simulated PL and PLE spectra consisting of zero-phonon line, and one-phonon and two-phonon (^) sidebands, PL and PLE are a mirror-symmetric relationship.

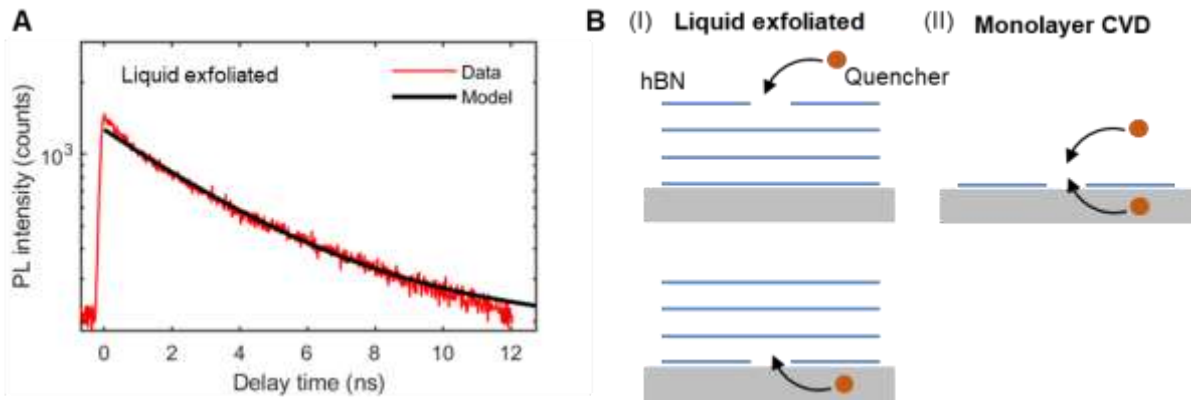

**Fig. S33.** (**A**) Time-resolved photoluminescence decay curve of liquid exfoliated hBN and a single exponential decay fit to determine photoluminescence lifetime as an indicator of how isolated away from quenching source. For instance, (**B**) (I) oxygen and moisture are likely to be a quencher to vacancy centre in a top layer; charge impurities in the $SiO_2$ substrate of hBN are likely to be a quencher to vacancy centre in a bottom layer, which cause photoluminescence lifetime to shorten and become significant in (II) monolayer CVD hBN.

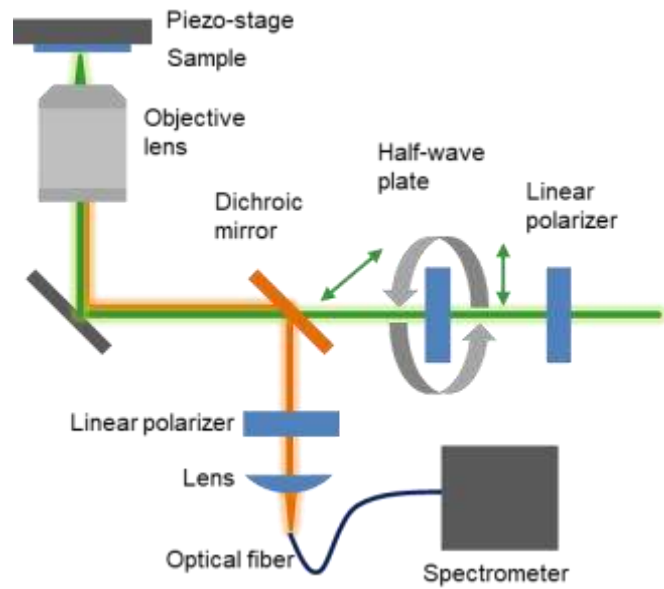

**Fig. S34.** Optical configuration for time-resolved photoluminescent polarization measurements of multilayer CVD hBN, where polarization of laser light is controlled by rotating half-wave plate.

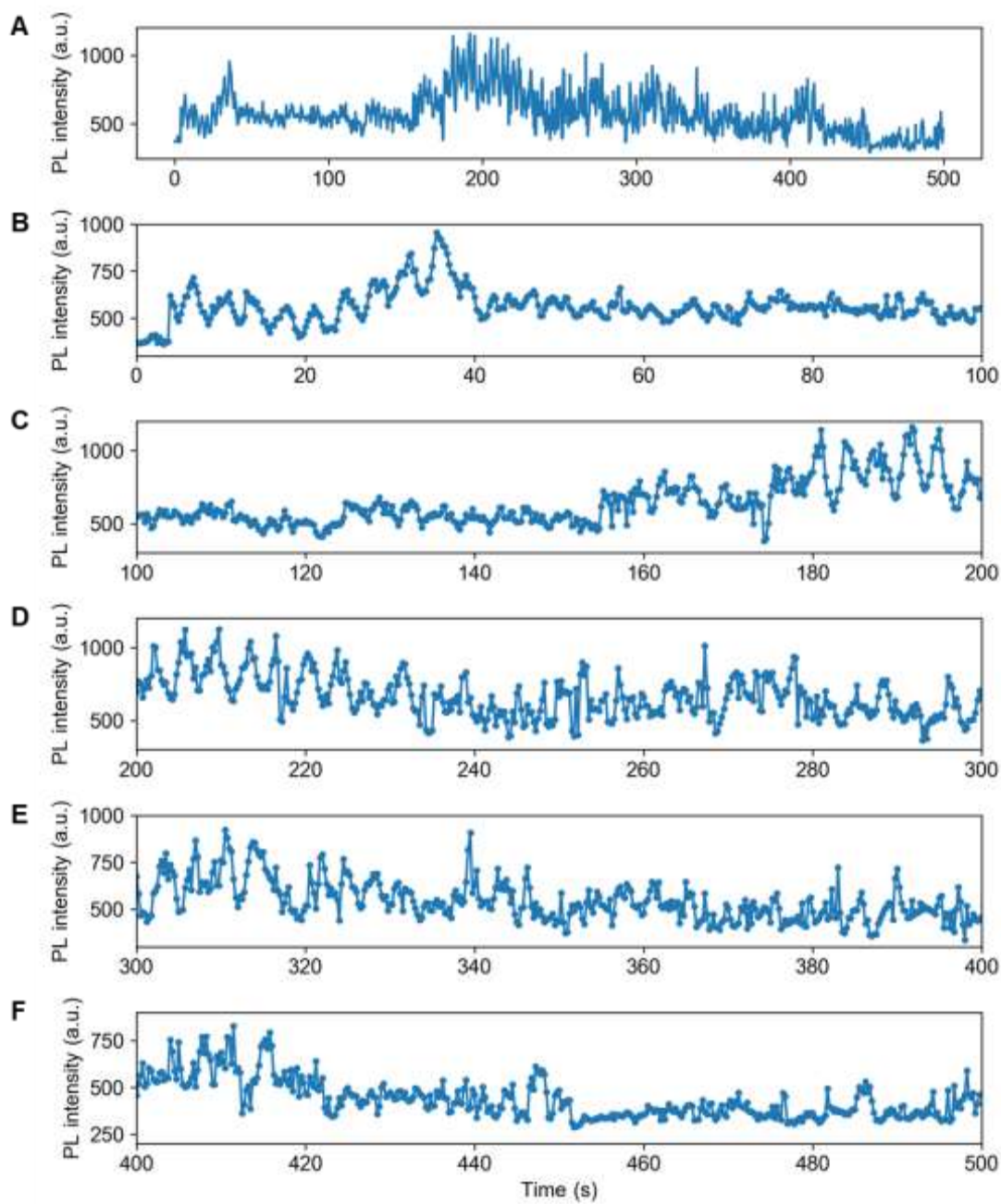

**Fig. S35. Exemplary raw data of the time-resolved polarization measurement**. (**A**) Whole and (**B**-**F**) 100 s-zoomed in of intensity time traces rotating linear polarization shown in Figure 4B and 4C, showing. The whole intensity time trace displays the overall trend of intensity change due to the blinking. The zoomed in images shows the periodic change in intensity along the continuous rotation of linearly polarized excitation and sudden change in the dipole angle.

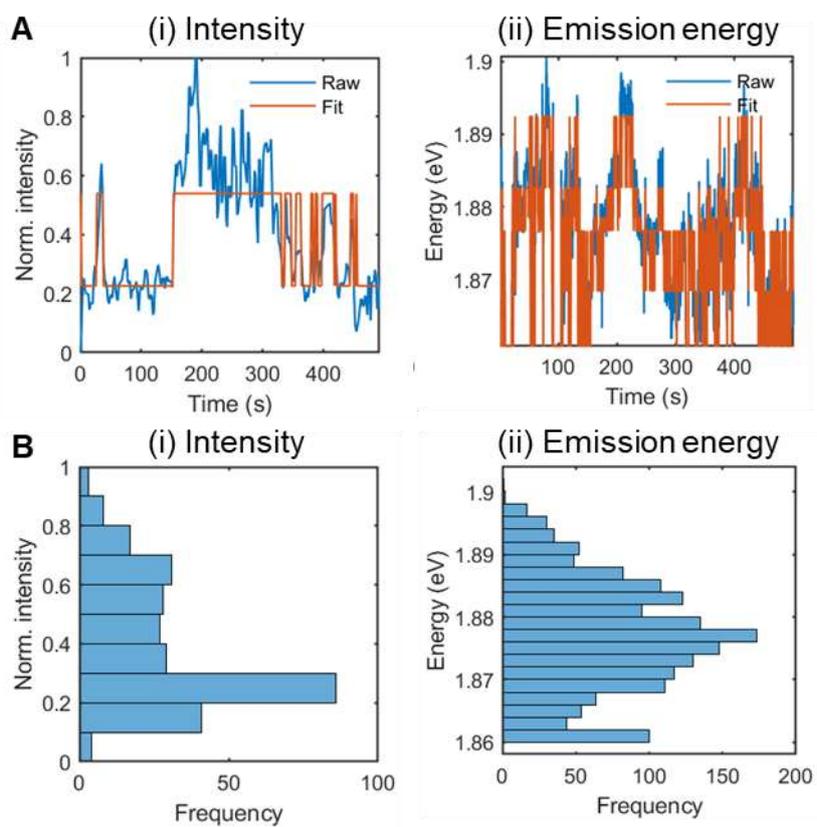

**Fig. S36.** (**A**) Time traces of intensity (i) and emission energy (ii). The traces were fitted with a cluster analysis algorithm to determine states. Histograms of the traces (**B**) were discrete.

**Fig. S 37.** Distributions of dwell time (red dots) of CVD multilayer hBN extracted from the time traces of (**A**) intensity (*I*), (**B**) phase shift (*φ*), and (**C**) emission energy (*E*) on the time-

resolved polarization measurements. The distribution showed single exponential decay and was fitted (black solid lines) to obtain dwell time constants.

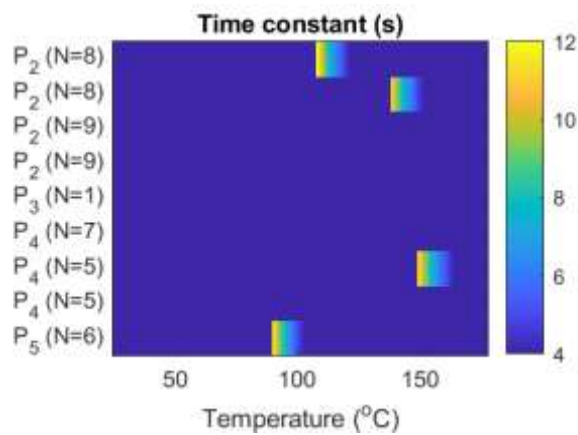

**Fig. S38.** Calculation of temperature dependent time constant for various vacancy structures based on the barrier energy calculated above with Eq. (S2). The colour of blue to yellow in the map corresponds to the time constant of 4-12 s obtained in the time-resolved polarization measurement.

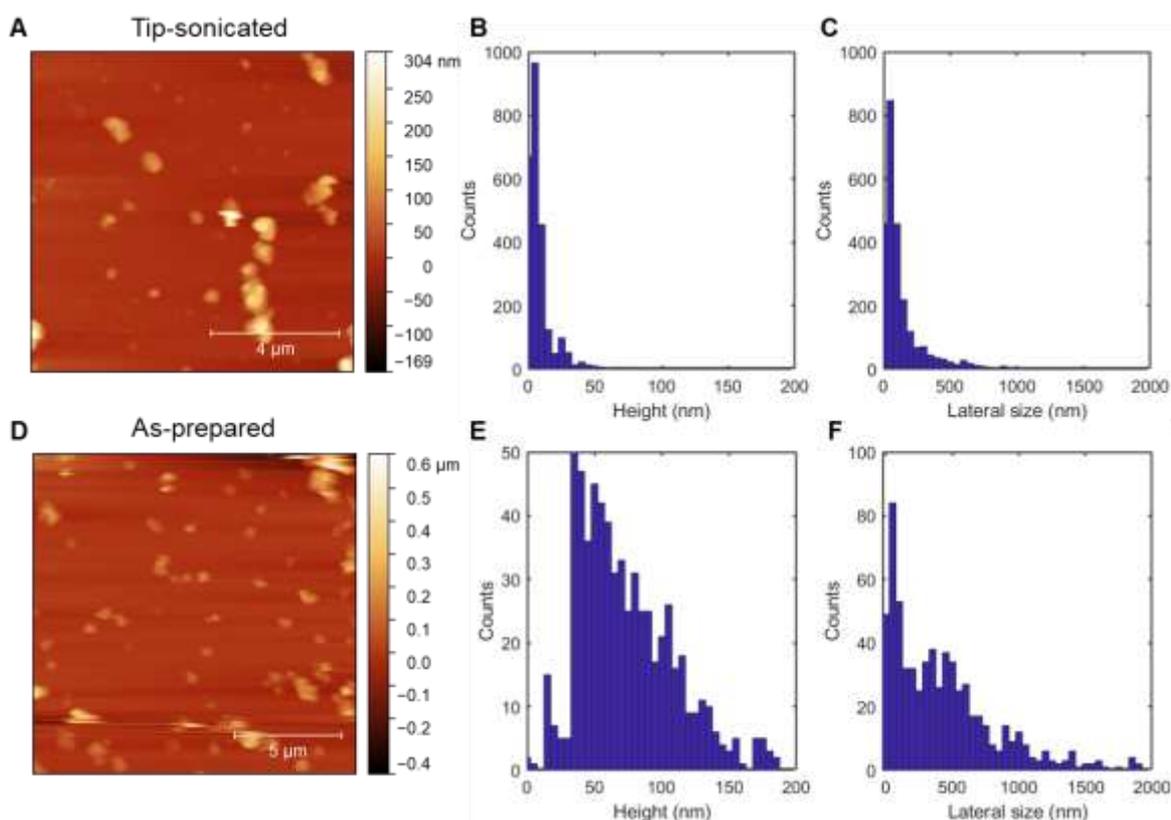

**Fig. S39.** (**A, D**) Atomic force micrograph, histograms of (**B, E**) thickness, and (**C, F**) lateral size of tip-sonicated and as-prepared hBN prepared by liquid-exfoliation, respectively, showing that the tip-sonication process produced thinner and smaller flakes. The median thickness and lateral size of the tip-sonicated hBN are 66 nm (10th and 90th percentiles: 39 and 129 nm) and 358 nm (58 and 1013 nm), respectively, whereas those of as-prepared hBN are 9 nm (3 and 25 nm) and 97 nm (48 and 376 nm), respectively.

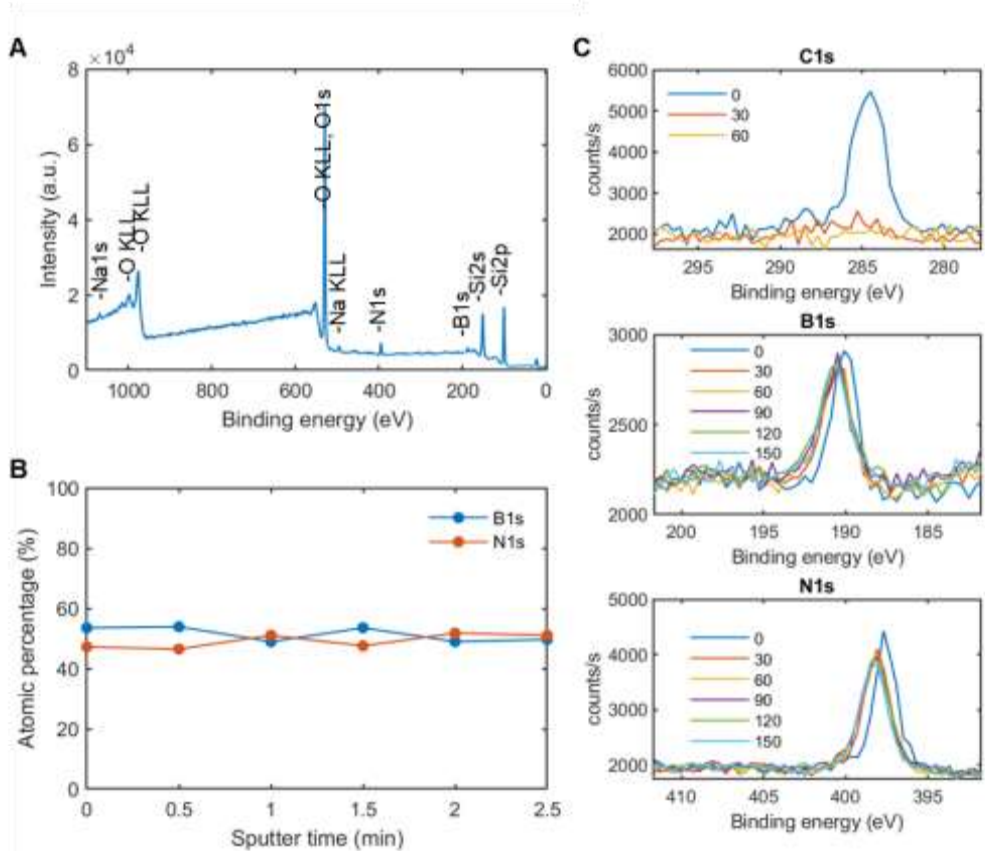

**Fig. S40.** XPS spectra of tip-sonicated hBN flakes after gradual ion sputtering (500eV, 500nA, 2x2 mm raster, 30 sec intervals). (A) XPS spectra after 1 min sputtering, showing reduced traces of carbon and potassium. (B) Atomic percentage analysis for B:N for different sputtering times. (C) XPS spectra for C1s, B1s and N1s peaks for different sputtering times show gradual decrease of C1s, but no change in B1s and N1s.

**Table S1.** Number of vacancy defects in a triangular, monoatomic 2D lattice, as a function of the size of the vacancy ($N$) [48].

| $N$ | Number of Vacancy Defects |
|---|---|
| **1** | 1 |
| **2** | 1 |
| **3** | 1 |
| **4** | 3 |
| **5** | 4 |
| **6** | 12 |
| **7** | 24 |
| **8** | 66 |
| **9** | 159 |
| **10** | 444 |
| **11** | 1161 |
| **12** | 3226 |
| **13** | 8785 |
| 14 | 24453 |
| 15 | 67716 |
| 16 | 189309 |

**Table S2.** Lifetimes of the most-probable nanopore isomers in hBN, for each size, $N$, of the nanopore considered in this work. The lifetime is calculated based on the atom type which is etched to form the next larger nanopore (B – boron, N – nitrogen, ZZ – zigzag, AC – armchair, SB – singly bonded), and the activation energy corresponding to the etching of that atom, at a temperature of 500 °C. The rows containing the nanopores having the largest lifetimes amongst all the nanopores considered are highlighted in orange colour, denoting the most-stable pores.

| $N$ | Atom Type to be Etched | Activation Energy (eV) | Lifetime (s) |
|---|---|---|---|
| 1 | NZZ | 2.303 | $1.03\times10^2$ |
| 2 | BZZ | 0.506 | $1.99\times10^{-10}$ |
| 3 | BZZ | 0.506 | $1.99\times10^{-10}$ |
| 4 | NZZ | 2.303 | $1.03\times10^2$ |
| 5 | BZZ | 0.506 | $1.99\times10^{-10}$ |
| 6 | BAC | 1.321 | $4.06\times10^{-5}$ |
| 7 | NSB | 1.141 | $2.75\times10^{-6}$ |
| 8 | BZZ | 0.506 | $1.99\times10^{-10}$ |
| 9 | NZZ | 2.303 | $1.03\times10^2$ |
| 10 | BZZ | 0.506 | $1.99\times10^{-10}$ |
| 11 | BAC | 1.321 | $4.06\times10^{-05}$ |
| 12 | NSB | 1.141 | $2.75\times10^{-6}$ |
| 13 | BAC | 1.321 | $4.06\times10^{-5}$ |
| 14 | NSB | 1.141 | $2.75\times10^{-6}$ |
| 15 | BZZ | 0.506 | $1.99\times10^{-10}$ |
| 16 | NZZ | 2.303 | $1.03\times10^2$ |

**Table S3.** Convergence of the plane wave cutoff, using a 1×1×1 Γ-centreed $k$-point mesh, for DFT calculations using a 9×9 monolayer hexagonal boron nitride supercell, using PAW pseudopotentials [13], and the Perdew-Burke-Ernzerhof (PBE) [14] generalized gradient approximation exchange-correlation functional. The system energy is converged to less than 2 meV/atom using a plane wave cutoff of 500 eV.

| Energy Cutoff (eV) | Energy of System (eV) | Difference (eV) | Difference (meV/atom) |
|---|---|---|---|
| 300 | -1434.5083 | 0.8093 | 4.9955 |
| 400 | -1435.3176 | 0.5674 | 3.5027 |
| 500 | -1435.8850 | 0.2703 | 1.6684 |
| 600 | -1436.1553 | | |

**Table S4.** Convergence of the Monkhorst-Pack-based [49] $k$-point mesh, using a plane wave cutoff of 500 eV and projector augmented wave (PAW) pseudopotentials [13], for density functional theory (DFT) calculations using a 9×9 hexagonal boron nitride supercell, and the Perdew-Burke-Ernzerhof (PBE) [14] generalized gradient approximation exchange-correlation functional. The system energy is converged to less than 0.01 meV/atom using a 1×1×1 $k$-point mesh, *i.e.*, Γ-point sampling.

| $k$-Point Mesh | Energy of System (eV) | Difference (eV) | Difference (meV/atom) |
| --- | --- | --- | --- |
| 1×1×1 | -1439.8167 | -0.0014 | -0.0088 |
| 2×2×1 | -1439.8153 | 3×10$^5$ | -0.0001 |
| 2×2×2 | -1439.8153 | | |

**Table S5. Determination of the Optimized Lattice Parameter.** Convergence of the Monkhorst-Pack-based [49] $k$-point mesh, using a plane wave cutoff of 500 eV, for DFT calculations using a the primitive hexagonal boron nitride (hBN) unit cell, using PAW pseudopotentials [13], and the Perdew-Burke-Ernzerhof (PBE) [14] generalized gradient approximation exchange-correlation functional. The system energy is converged to less than 0.0002 meV/atom using an 8×8×2 $k$-point mesh. Using an 8×8×2 $k$-point mesh, a plane wave cutoff of 500 eV, and Grimme's D3 method [15] to account for dispersion interactions, we optimized the hBN unit cell, using the PBE exchange-correlation functional, till the maximum force on any atom was less than 0.005 eV/Å. We found the optimized B-N bond length to be 1.447 Å, which compares very well with the experimental value [50] of 1.446 Å (within 0.07%)!

| $k$-Point Mesh | Energy of System (eV) | Difference (eV) | Difference (meV/atom) |
| --- | --- | --- | --- |
| 4×4×1 | -35.6238 | -0.0892 | -0.0223 |
| 8×8×2 | -35.7129 | 0.0009 | -0.0002 |
| 16×16×4 | -35.7120 | | |

**Table S6.** Calculated energy gap of the initial (N) and final (N*) states for N = 1, 5, 6, 7, and 8 of lattice vacancies in monolayer hBN.

| Pore N | ($N_B$, $N_N$) | HSE 06 Gap (eV) |
| --- | --- | --- |
| 1 | (0,1) | 2.11 |
| 1* | (0,1) | 2.89 |
| 1 | (1,0) | 1.37 |
| 5 | (3,2) | 1.95 |
| 5* | (3,2) | 1.62 |
| 6 | (4,2) | 1.84 |
| 6* | (4,2) | 1.52 |
| 7 | (5,2) | 1.94 |
| 7* | (5,2) | 2.22 |
| 8 | (5,3) | 2.30 |
| 8* | (5,3) | 1.46 |
| 8*$^2$ | (5,3) | 0.92 |

**Table S7.** The size of defects, calculated energy gap of the most likely isomers based on HSE06 and corresponding assignment (i).

| $N$ | HSE06 Gap (eV) | Experiment (eV) | Assignment |
|---|---|---|---|
| 4 | 4.4289 | | - |
| 2 | 4.3163 | | - |
| 10 | 2.404 | 2.26±0.013 | $P_1$ |
| 8 | 2.2981 | 2.15±0.030 | $P_2$ |
| **16** | 2.2779 | 2.15±0.030 | $P_2$ |
| 14 | 2.2669 | 2.15±0.030 | $P_2$ |
| **9** | 2.2214 | 2.15±0.030 | $P_2$ |
| 15 | 2.188 | 2.15±0.030 | $P_2$ |
| **1** | 2.1135 | 1.98±0.017 | $P_3$ |
| 5 | 1.9508 | 1.90±0.021 | $P_4$ |
| 7 | 1.9459 | 1.90±0.021 | $P_4$ |
| 6 | 1.8382 | 1.80±0.028 | $P_5$ |
| 12 | 1.6327 | 1.70±0.037 | $P_6$ |
| 13 | 1.3698 | | |
| 3 | 0.8732 | | |
| 11 | 0.8365 | | |

**Table S8.** The number of defects for monolayer and bilayer CVD hBN observed in the STEM imaging.

| Monolayer | | | Bilayer | | |
|---|---|---|---|---|---|
| $N$ | Number of observed defects | Assignment | $N$ | Number of observed defects | Assignment |
| 1 | 8 | $P_3$ | 1 | 1 | $P_3$ |
| 4 | 3 | - | 4 | 3 | - |
| 9 | 1 | $P_2$ | 13 | 1 | - |
| | | | 16 | 4 | $P_2$ |

**Table S9**. Extracted fitting parameters from the boric acid etching experiments of liquid exfoliated hBN.

| Fitting Parameters | Extracted Values |
|---|---|
| $k_a$ | 0.00288 1/min (850 °C) |
| | 1.22×10$^{-19}$ 1/min (220 °C) |
| $k_b$ | 0.125 1/min (850 °C) |
| | 8.83×10$^{-15}$ 1/min (220 °C) |
| $k_c$ | 0.251 1/min (850 °C) |
| | 0.0241 1/min (220 °C) |
| $C_1(0)$ | 6.71×10$^{-2}$ 1/um$^2$ |
| $k_{a0}$ | 1.88×10$^{10}$ 1/min |
| $k_{b0}$ | 2.46×10$^9$ 1/min |
| $E_a$ | 2.86 eV |
| $E_b$ | 2.30 eV |

**Table S10.** Photon energy of zero-phonon line (ZPL), the first phonon side band (PSB), and the energy difference between the zero-phonon line and phonon side band (Δ$E$) extracted from photoluminescence (PL) and excitation (PLE) spectra shown in **Fig. 4D**.

| Assignment | PL | | | | PLE | |
|---|---|---|---|---|---|---|
| | ZPL (eV) | PSB (eV) | Δ$E$ (eV) | Wavenumber (cm$^{-1}$) | PSB (eV) | Δ$E$ (eV) |
| P$_1$ | 2.202 | 2.036 | 0.166 | 1339 | 2.371 | 0.169 |
| P$_2$ | 2.189 | 2.023 | 0.166 | 1339 | 2.348 | 0.159 |
| P$_3$ | 1.964 | 1.806 | 0.158 | 1274 | 2.149 | 0.185 |
| P$_4$ | 1.900 | 1.738 | 0.162 | 1307 | 2.036 | 0.136 |
| P$_5$ | 1.856 | 1.697 | 0.159 | 1282 | - | - |
| P$_6$ | 1.711 | - | - | - | - | - |

**Table S11.** Dwell time constants for individual assigned peaks of intensity, phase, and energy time traces obtained from the time-resolved polarization measurements. The constants were irrelevant to the assignments from P$_1$ through P$_6$.

| | Dwell time constant (s) | | | | | |
|---|---|---|---|---|---|---|
| | P$_6$ | P$_5$ | P$_4$ | P$_3$ | P$_2$ | P$_1$ |
| **Intensity** | 4.0 | 7.2 | 10.5 | 6.5 | 7.9 | 6.4 |
| **Phase** | 11.8 | 11.3 | 10.7 | 10.4 | 9.2 | 7.8 |
| **Energy** | 4.1 | 6.0 | 5.4 | 5.6 | 5.5 | 4.9 |